\documentclass[aps,prd,a4paper,twocolumn,amsmath,showpacs,superscriptaddress,nofootinbib,preprintnumbers]{revtex4-1}
\usepackage{verbatim}
\usepackage[T1]{fontenc}
\usepackage[utf8]{inputenc}
\usepackage[american]{babel}
\usepackage{epsfig}
\usepackage{mathrsfs}
\usepackage{graphicx}
\usepackage{booktabs}
\usepackage{multirow}
\usepackage{dcolumn}
\usepackage{amsmath}
\usepackage{subcaption}
\usepackage{mathtools}
\usepackage{amsfonts}
\usepackage{amssymb}
\usepackage{epstopdf}
\usepackage{bm}
\usepackage{siunitx}
\usepackage{braket}
\usepackage{enumitem}
\usepackage{soul}
\usepackage{fontawesome}
\usepackage[table]{xcolor}
\usepackage{color}
\usepackage{transparent}
\usepackage{pifont}
\definecolor{navyblue}{rgb}{0.0, 0.0, 0.5}
\definecolor{royalblue}{rgb}{0.25, 0.41, 0.88}
\definecolor{cadmiumgreen}{rgb}{0.0, 0.42, 0.24}
\definecolor{blue-violet}{rgb}{0.54, 0.17, 0.89}
\definecolor{darkviolet}{rgb}{0.58, 0.0, 0.83}
\definecolor{orange(colorwheel)}{rgb}{1.0, 0.5, 0.0}

\usepackage{hyperref}
\hypersetup{
    colorlinks=true, 
    linkcolor=royalblue, 
    citecolor=magenta}

\newcommand\ee{\end{equation}}
\newcommand\be{\begin{equation}}
\newcommand\eea{\end{eqnarray}}
\newcommand\bea{\begin{eqnarray}}

\newcommand\mev{\mathrm{MeV}}

\renewcommand\[{\left[}

\newcommand{\neff}{N_{\rm eff}}

\newcommand{\lcdm}{\Lambda\mathrm{CDM}}
\newcommand{\qdmap}{Q_{\mathrm{DMAP}}}

\newcommand\eq[1]{Eq.~\eqref{eq:#1}}
\newcommand\sect[1]{Sec.~\ref{sec:#1}}
\newcommand\fig[1]{Fig.~\ref{fig:#1}}
\newcommand\tab[1]{Tab.~\ref{tab:#1}}

\newcommand{\mscr}[1]{\mathscr{#1}}

\newcommand{\mcal}[1]{\mathcal{#1}}

\usepackage{booktabs}
\usepackage{multirow}
\usepackage{dcolumn}
\usepackage{colortbl}

\definecolor{magenta(process)}{rgb}{1.0, 0.0, 0.56}

\definecolor{darkspringgreen}{rgb}{0.09, 0.45, 0.27}

\definecolor{royalblue(web)}{rgb}{0.25, 0.41, 0.88}
\definecolor{Gray}{gray}{0.9}
\definecolor{softred}{rgb}{0.8, 0.3, 0.3} 
\definecolor{softgreen}{rgb}{0.3, 0.7, 0.3}
\definecolor{softerred}{rgb}{0.94, 0.5, 0.5}   
\definecolor{softergreen}{rgb}{0.5, 0.88, 0.5}  
\usepackage{todonotes}

\begin{document}

\title{One Extension to Explain Them All, One Scale-Invariant Spectrum to Test Them All, and in One Model Bind Them}

\author{Matteo Forconi}
\email{matteo.forconi@fe.infn.it}
\affiliation{Physics Department and INFN sezione di Ferrara,  Università degli Studi di Ferrara, via Saragat 1, I-44122 Ferrara, Italy}
\affiliation{School of Mathematical and Physical Sciences, University of Sheffield, Hounsfield Road, Sheffield S3 7RH, United Kingdom}

\author{Eleonora Di Valentino}
\email{e.divalentino@sheffield.ac.uk}
\affiliation{School of Mathematical and Physical Sciences, University of Sheffield, Hounsfield Road, Sheffield S3 7RH, United Kingdom}

\date{\today}

\preprint{}
\begin{abstract}

The increasing precision of Cosmic Microwave Background (CMB) observations has unveiled significant tensions between different datasets, notably between Planck and the Atacama Cosmology Telescope (ACT), as well as with late-Universe measurements of the Hubble constant. In this work, we explore a variety of beyond-$\Lambda$CDM extensions to assess their ability to reconcile these discrepancies. Specifically, we consider modifications to the primordial power spectrum, geometry, dark energy, the effective number of relativistic species, and the primordial helium fraction, as well as an Early Dark Energy (EDE) component. We evaluate each model using multiple statistical tools, including the Akaike Information Criterion (AIC), Bayesian model comparison, suspiciousness, and the goodness-of-fit estimator. 
Our results confirm that no single extension fully resolves all existing tensions. While a nonzero curvature is favored by Planck-only data, it does not alleviate the Planck-ACT discrepancy. The EDE scenario, particularly with a fixed Harrison-Zeldovich spectrum, provides the best resolution to the Planck-ACT inconsistency, while $w$CDM is more effective at reducing the Hubble tension when SH0ES data are included. However, the statistical preference for these extensions remains moderate, and imposing $n_s=1$ often worsens model performance. Our findings highlight the limitations of modifications to $\Lambda$CDM and suggest that either more complex new physics or, more likely, improved systematic understanding in the CMB sector may be required to fully address the observed tensions. While CMB experiments are often considered the gold standard of precision cosmology, our results reinforce that these measurements are not immune to systematic uncertainties, which may be underestimated in current analyses.

\end{abstract}

\maketitle

\section{Introduction}

The Cosmic Microwave Background (CMB) provides one of the strongest observational pillars of modern cosmology. Ever since the COBE satellite’s first measurements of the CMB’s nearly perfect black-body spectrum and tiny temperature fluctuations~\cite{1994ApJ...420..445F,Bennett:1996ce}, researchers have steadily refined these observations through missions such as WMAP~\cite{2013ApJS..208...19H,2013ApJS..208...20B} and, more recently, Planck~\cite{Planck:2018vyg}, the Atacama Cosmology Telescope (ACT)~\cite{ACT:2020frw,ACT:2020gnv}, and the South Pole Telescope (SPT)~\cite{SPT-3G:2014dbx,SPT-3G:2021eoc,SPT-3G:2021wgf}. These data sets allow the construction of a standard model of cosmology: $\Lambda$CDM, in which a spatially flat Universe is dominated at late times by a cosmological constant, $\Lambda$, and cold dark matter (CDM). Furthermore, in the early epochs, the Universe experienced a phase of exponential acceleration, the so-called inflationary epoch~\cite{Guth:1980zm}, which is responsible for large-scale homogeneity and the generation of primordial density fluctuations. However, it is important to underline that $\Lambda$CDM remains a phenomenological framework containing three major unknown ingredients—inflation, dark matter, and dark energy—whose theoretical motivations still lack direct experimental confirmation.

Over the last decade, as the precision of cosmological measurements has improved, several intriguing discrepancies have emerged~\cite{Abdalla:2022yfr,DiValentino:2022fjm,2022NewAR..9501659P}. The most significant is the so-called Hubble tension~\cite{Riess:2019qba,DiValentino:2020vnx,Staicova:2021ntm,Vagnozzi:2023nrq,Verde:2019ivm,DiValentino:2020zio,DiValentino:2021izs,Schoneberg:2021qvd,Shah:2021onj,Kamionkowski:2022pkx,Giare:2023xoc,Hu:2023jqc,Verde:2023lmm,Benisty:2024lmj,DiValentino:2024yew,Perivolaropoulos:2024yxv,Erdem:2024vsr}, a $>5\sigma$ mismatch between the Hubble constant measured via late-Universe probes~\cite{Freedman:2020dne,Birrer:2020tax,Wu:2021jyk,Anderson:2023aga,Scolnic:2023mrv,Jones:2022mvo,Anand:2021sum,Freedman:2021ahq,Uddin:2023iob,Huang:2023frr,Li:2024yoe,Pesce:2020xfe,Kourkchi:2020iyz,Schombert:2020pxm,Blakeslee:2021rqi,deJaeger:2022lit,Murakami:2023xuy,Breuval:2024lsv,Freedman:2024eph,Riess:2024vfa,Vogl:2024bum,Gao:2024kkx,Scolnic:2024hbh,Said:2024pwm,Boubel:2024cqw} and the smaller $H_0$ value inferred from early-Universe observations~\cite{Planck:2018vyg,ACT:2020gnv}, under the assumption of $\Lambda$CDM. Moreover, Planck data itself exhibit other anomalies, including an enhanced CMB lensing amplitude at about $2.8\sigma$~\cite{Planck:2013pxb, DiValentino:2013mt, Planck:2015fie, Addison:2015wyg, Planck:2016tof, Renzi:2017cbg, DiValentino:2017rcr, Planck:2018vyg, Addison:2023fqc,Ben-Dayan:2024uvx,Specogna:2024euz,Rosenberg:2022sdy} and a preference for a closed geometry at roughly $3.4\sigma$~\cite{Planck:2018vyg}, introducing an inconsistency with Baryon Acoustic Oscillation (BAO) data~\cite{Handley:2019tkm,DiValentino:2020srs,Vagnozzi:2020rcz,Shumaylov:2021qje,Yang:2022kho,Glanville:2022xes,DiValentino:2020hov,Anselmi:2022uvj}. Collectively, such anomalies raise the question of whether new physics beyond $\Lambda$CDM is needed or whether there are some unrecognized systematic errors in one or more datasets.

An additional layer of tensions arises from comparing multiple independent CMB data sets. While the Planck satellite has provided the tightest constraints to date, ACT can offer valuable confirmation of Planck cosmology. Intriguingly, although ACT generally agrees with Planck, small discrepancies appear. For instance, analyses have found a mild-to-moderate tension at the $\sim2.5\sigma$ level between Planck and ACT~\cite{Lin:2019zdn,Handley:2020hdp,DiValentino:2022rdg,LaPosta:2022llv,Calderon:2023obf}, particularly in the inferred values of the spectral tilt $n_s$~\cite{Jiang:2022uyg,DiValentino:2018zjj,Ye:2022efx,Jiang:2022qlj,Giare:2023wzl}, but also in the effective number of relativistic degrees of freedom~\cite{DiValentino:2022oon}, the running and the running of the running of the spectral index~\cite{ACT:2020gnv,Forconi:2021que}, the neutrino mass~\cite{DiValentino:2021imh}, the evidence for Early Dark Energy~\cite{Hill:2021yec,Poulin:2021bjr}, and for Dynamical Dark Energy~\cite{Giare:2024ocw}.

These developments underline the importance of assessing the consistency of independent CMB measurements~\cite{Charnock:2017vcd,Lin:2019zdn,Handley:2019wlz,Hazra:2024nav} and their compatibility with external late-Universe observations such as SH0ES. In this work, we analyze different extensions to the standard cosmological scenarios using multiple statistical metrics. We investigate how these models perform in light of the Hubble tension and the Planck vs. ACT discrepancy, aiming to find a unique solution. 

This paper is organized as follows. \sect{BG} is dedicated to presenting the alternative cosmological models. \sect{method} describes the statistical metrics used in the analyses. \sect{MD} presents the methodology and the dataset, whereas in \sect{results}, we present our results. We conclude in \sect{concl}.

\section{Battle Grounds}
\label{sec:BG}

In this paper, we aim to study the consistency between Planck and ACT observations within different scenarios. Before proceeding to analyze the different extra parameters we choose to employ, it is important to stress that there are mainly two indicators of the discrepancy between these two CMB experiments: the spectral index, $n^{\rm Planck}_s=0.9649\pm0.0044$~\cite{Planck:2018vyg} vs. $n^{\rm ACT}_s=1.008\pm0.015$~\cite{ACT:2020gnv}, and the baryon energy density, $\Omega^{\rm Planck}_bh^2=0.02236\pm0.00015$~\cite{Planck:2018vyg} vs. $\Omega^{\rm ACT}_bh^2=0.02153\pm0.00030$~\cite{ACT:2020gnv}. These differences persist even when accounting for the lack of small-multipole ($\sim$ first two acoustic peaks) observations from ACT~\cite{ACT:2020gnv,Giare:2022rvg} or its limited amount of polarization data (only for $n_s$)~\cite{Giare:2022rvg}, as well as when including non-CMB data~\cite{ACT:2020gnv,Giare:2022rvg}. 
Although observational systematic errors and statistical fluctuations may provide an explanation, we aim to explore the possibility that a limitation of the standard cosmological model is the root of this tension.

The idea behind choosing the alternative models is that they should impact either $n_s$ or $\Omega_bh^2$, such that a natural shift occurs that might ease the tension. Furthermore, to better study the behavior of the consistency between Planck and ACT, quantified by the statistical tools described in \sect{method}, we also force $n_s=1$ for each case, as explained below. By reducing the number of parameters in the model, we accommodate ACT predictions for the spectral index and test the ability of Planck observations to compensate. On the other hand, we also move in the opposite direction (towards Planck predictions) by imposing a tight bound on the helium fraction, forcing higher values of $\Omega_bh^2$ with respect to ACT observations in one realization.

The proposed \textit{battle grounds} to study this inconsistency range from early- to late-Universe solutions. This is because Planck shows a moderate deviation from $\Lambda$CDM when extending the latter, while in contrast, ACT shows anomalies in the former~\cite{Giare:2023xoc}. Although the large experimental uncertainties in the late-time sector suggest that the discrepancy between the two probes could stem from an inconsistency in the early Universe, we include late-time solutions for two main reasons: first, to explore the possibility that this anomaly is somehow linked to the Hubble tension and determine whether any of the beyond-$\Lambda$CDM phenomenology studied can increase the present-day expansion rate of the Universe while also leading to an agreement between the two CMB probes; and second, because for each model, we also impose $n_s=1$, thereby introducing a modification at early times.

\subsubsection{Primordial Power Spectrum}

A general prediction of inflationary theory is that the power spectrum of density perturbations can be well described by a power law $\propto k^{n_s}$, where $n_s$ is the scalar spectral index. The value of the spectral index should be nearly (but not exactly) one due to the dynamics of the inflaton field. This contrasts with the phenomenological model of the early Universe that fixes the scalar spectral index to $n_s=1$, the so-called Harrison-Zeldovich (HZ) spectrum~\cite{Harrison:1969fb,Zeldovich:1972zz,Peebles:1970ag}. 
Although it is possible to recover the HZ spectrum in specific inflationary models (e.g.,~\cite{Vallinotto:2003vf,Rinaldi:2015uvu,Cecchini:2024xoq,Rinaldi:2023mdf,Fu:2023tfo}), generally, a value of $n_s \neq 1$ is considered strong evidence in favor of inflationary theory. For example, tensor modes in the CMB are a unique prediction of inflationary theory, and their amplitude is expected to be proportional to the square of the deviation from scale invariance, $\lvert n_s - 1 \rvert^2$. Furthermore, as shown in~\cite{Ye:2021nej,Ye:2022efx,Jiang:2022uyg,Giare:2024akf}, a shift in $n_s$ is related to a shift in $H_0$ as $\delta n_s \simeq 0.4 \delta H_0 / H_0$, assuming a non-modified recombination process and a late-time $\Lambda$CDM framework. From this perspective, we can argue that an HZ spectrum could provide a solution to the Hubble tension.

The latest CMB observations provided by \textit{Planck} exclude an HZ spectrum at more than $8\sigma$~\cite{Planck:2018nkj,Planck:2018vyg,Planck:2018jri}, corroborating inflationary theory. On the other hand, the ACT collaboration shows deviations from the $\Lambda$CDM scenario~\cite{DiValentino:2022rdg,DiValentino:2022oon,Calderon:2023obf}, including a discrepancy in the scalar spectral index, which is consistent with $1$ within one $\sigma$ ($n_s=1.008\pm0.015$~\cite{ACT:2020gnv}). 
This represents a new potential challenge for inflationary cosmology: is canonical inflation the dominant mechanism for producing the perturbations in the early Universe? The role of this new tension in the spectral index could be related to systematic effects or could play a prominent role in our understanding of the Universe~\cite{Giare:2022rvg,DiValentino:2018zjj,Ye:2022efx,Jiang:2022qlj,Jiang:2022uyg,Benetti:2017juy,Benetti:2017gvm,Benetti:2013wla,Pandolfi:2010dz}.

In the standard $\Lambda$CDM model, the spectral index is assumed to be scale-independent. This means neglecting the next-order correction terms from the Taylor expansion around the pivot scale $k_\star$ and considering the spectrum of primordial perturbations as the power law:
\begin{equation}
    \Delta^2(k)=A_s\left(\frac{k}{k_\star}\right)^{n_s-1}\,.
    \label{eq:powerlaw}
\end{equation}
However, inflation does not predict a precise value for the spectral index. This depends on the details of the inflationary dynamics and the shape of the potential. For this reason, we can expand \eq{powerlaw} to the next order:
\begin{multline}
    \Delta(k)=\Delta(k_\star)+\left.\frac{d\Delta(k)}{d\log{k}}\right\rvert_{k_\star}\log{\left(\frac{k}{k_\star}\right)}\\
    +\frac12\left.\frac{d^2\Delta(k)}{d\log{k}^2}\right\rvert_{k_\star}\log^2{\left(\frac{k}{k_\star}\right)}+...
\end{multline}
and define the running of the spectral index $\alpha_s$~\cite{Kosowsky:1995aa,Lidsey:1995np,Kuroyanagi:2008ye,Kuroyanagi:2011iw,Zarei:2014bta}\footnote{Similarly, you can define it for the tensor index, see, e.g., ~\cite{Giare:2020vhn,Giare:2020vss}} as
\begin{equation}
    \alpha_s\equiv\left.\frac{dn_s}{d\ln{k}}\right\rvert_{k_\star},
    \label{eq:running}
\end{equation}
which quantifies the rate of change of $n_s$ per Hubble time. Planck observations~\cite{Planck:2018vyg} provide a running compatible with zero, $\alpha_s=-0.006\pm0.013$ at $68\%$ CL, which maintains its compatibility when combined with other probes~\cite{Li:2018iwg,Forconi:2021que,Giare:2019snj,Martin:2024qnn,Martin:2024nlo,Ballardini:2024irx}. The constraints are slightly relaxed for ACT and SPT-3G~\cite{2011ApJ...739...52D,2011ApJ...743...28K,Forconi:2021que,DiValentino:2022rdg,Giare:2022rvg}, but still favor a scale-independent spectral index at $2\sigma$, as also suggested by forecasts of future experiments~\cite{Bahr-Kalus:2022prj,Easther:2021rdg}.

\subsubsection{Late Dark Energy}

Since our Universe is accelerating~\cite{SupernovaSearchTeam:1998fmf,SupernovaCosmologyProject:1998vns,2011PhRvL.107b1302S,Moresco:2016mzx,Haridasu:2017lma}, the dominant component of our Universe today must be characterized by a negative pressure, so that it can counterbalance the natural decelerated expansion. This component is called Dark Energy ($\sim 70\%$ of the total content), and in the standard model of cosmology, it is parametrized by a positive cosmological constant, $\Lambda$. $\Lambda$ has an equation of state:
\begin{equation}
    w=\frac{p}{\rho}=-1
    \label{eq:EoS}
\end{equation} 
which satisfies the negative-pressure requirement ($w<-1/3$). 

An alternative to the cosmological constant $\Lambda$ is to relax \eq{EoS} and assume a generic fluid with a constant equation of state $w$. By extending $\Lambda$CDM to $w$CDM, that is, introducing $w$ as an extra parameter, it is possible to test the validity of $\Lambda$CDM. With $w$ now a free parameter, aside from recovering $\Lambda$CDM when $w=-1$, we can explore two different regimes: the \textit{phantom regime}~\cite{Caldwell:2003vq,Nojiri:2005sx,Frampton:2011sp,Astashenok:2012tv,Odintsov:2015zza,Odintsov:2018zai}, characterized by $w<-1$ and a possible violation of the null energy condition (although it is possible for the equation of state to be effectively phantom without violating the null energy condition), and the \textit{quintessence regime}~\cite{Wetterich:1987fm,Ratra:1987rm,Wetterich:1994bg,Caldwell:1997ii,Sahni:2002kh}. 
Planck data suggest a DE component in the phantom regime at $\lesssim2\sigma$, with $w=-1.58^{+0.52}_{-0.41}$ at $95\%$ CL~\cite{Planck:2018vyg}, and this preference is maintained when combined with other probes~\cite{Moresco:2016nqq,eBOSS:2020yzd,DAmico:2020ods,Grillo:2020yvj,Cao:2021irf,Chudaykin:2020ghx,Vagnozzi:2020dfn,Bargiacchi:2021hdp,Moresco:2022phi,Brieden:2022heh,Carrilho:2022mon,Semenaite:2022unt,KiDS:2020ghu,DES:2022ccp,DiValentino:2020evt}, while ACT is consistent with $w=-1$ but with large uncertainties ($w=-1.18^{+0.40}_{-0.55}$ at $68\%$ CL)~\cite{DiValentino:2022rdg,Giare:2023xoc}.

\subsubsection{Universe Geometry}

Spatial curvature is usually parametrized by the curvature energy density parameter $\Omega_k$, which affects the Hubble expansion rate with a redshift dependence $\propto (1+z)^2$. In modern cosmology, it has become common practice among phenomenologists and observers to consider the curvature parameter as optional, setting $\Omega_k=0$ and reducing to the standard six-parameter $\Lambda$CDM concordance cosmology. This approach is strongly motivated by inflationary theory~\cite{Guth:1980zm,Linde:1981mu,Linde:1983gd}. An exponential expansion during an inflationary phase drives the Universe towards flatness; in this phase, $\Omega_k=0$ is a late-time attractor and suppresses any primordial spatial curvature~\cite{Guth:1980zm}. However, fixing this parameter is akin to assuming a delta-function prior on a phenomenological parameter, which consequently alters the cosmological constraints. For this reason, it is important to have a solid justification for it~\cite{Anselmi:2022uvj}.

Curvature constraints from SPT-3G 2018 alone~\cite{SPT-3G:2021wgf} and ACT+WMAP~\cite{ACT:2020gnv} are fully consistent with a flat Universe. However, Planck data alone suggest a possible detection of a closed Universe (i.e., $\Omega_k<0$) at over $2\sigma$~\cite{Planck:2018vyg}. To break the geometric degeneracy~\cite{Bond:1997wr} that CMB experiments face when exploring curvature, temperature and polarization data are combined with late-time measurements such as Baryon Acoustic Oscillations (BAO)~\cite{Ross:2014qpa,BOSS:2016wmc}. This joint analysis reduces (and often eliminates) the preference for a negative curvature parameter, at the cost of combining experiments that are in tension at more than $3\sigma$ (see e.g.,~\cite{Handley:2019tkm,DiValentino:2020hov,DiValentino:2019qzk}). Some argue that the indication of a closed Universe arises from a prior problem in the exploration of parameter space with curvature~\cite{Efstathiou:2019mdh,Efstathiou:2020wem}. Conversely, the assumption of a flat Universe is reinforced when the HiLLiPoP likelihood is used~\cite{Tristram:2023haj}. 

Curvature can also be assessed through its effects on the late-time Universe alone. When flatness is not assumed in the fiducial model, CMB-independent analyses of low-redshift data do not provide compelling evidence for either a closed or flat Universe~\cite{Glanville:2022xes,Gonzalez:2021ojp,Favale:2023lnp,Bel:2022iuf,Yang:2022kho}. However, future analyses employing model-independent approaches may help resolve this conundrum~\cite{Amendola:2024gkz}.

\subsubsection{Helium Abundance}
\label{sec:Yp}

As fixing the spectral index exacerbates the tension in the baryon fraction of the Universe, we aim to test the tension between the two datasets by allowing the Helium fraction to vary freely. First, we simply apply a flat prior and refer to this extension as $\Lambda$CDM$+Y_p$. Then, we impose a prior on the primordial Helium mass fraction based on $Y_p=0.2446\pm0.0029$~\cite{Peimbert:2016bdg}, derived from H II region observations, which is similar to another widely used independent measurement, $Y_p=0.2449\pm0.0040$~\cite{Aver:2015iza};\footnote{We opt for the first one simply because we wanted to impose the most stringent limit on the helium mass fraction to observe a shift in $\Omega_b$.} we denote this model as $\Lambda$CDM$+$BBN. 

Introducing this parameter breaks the BBN relation~\cite{Serpico:2004gx}:
\begin{equation}
    \Omega_bh^2=\frac{1-0.007125Y_p^{\rm BBN}}{273.279}\left(\frac{T_{\rm CMB}}{2.7255K}\right)^3\eta_{10}
\end{equation}
where $\eta_{10}\equiv 10^{10}n_b/n_\gamma$ is the photon-baryon ratio today, $T_{\rm CMB}$ is the CMB temperature at the present time, and $Y_p^{\rm BBN}\equiv 4n_{\rm He}/n_b$ is the helium \textit{nucleon fraction}. The helium \textit{mass fraction} $Y_p$, used in our extensions, is recovered via the relation:
\begin{equation}
    Y_p=\frac{Y_{He4}m_{He4}}{Y_{He4}m_{He4}+(1-4Y_{He4})m_{H1}}
\end{equation}
with $Y_{He4}=1/4Y_p^{\rm BBN}$, $m_{He4}\simeq 4.0026$, and $m_{H1}\simeq 1.0078$.

\subsubsection{Effective Number of Relativistic Degrees of Freedom}

With the expansion of the Universe, the temperature drops below the electron mass ($T\sim 0.5\,\mev$), initiating pair annihilation between electrons and positrons. This results in the production of two photons with a subsequent injection of energy. Such an increase in energy does not affect neutrinos, as they decouple at $T_\nu\sim 1\,\mev$~\cite{Dolgov:2002ab,Wong:2002fa,Abazajian:2002qx} and therefore no longer share the same temperature, with $T_\nu\neq T_\gamma$. Specifically, they are related via
\begin{equation}
    T_\nu=\left(\frac{4}{11}\right)^{\frac{1}{3}} T_\gamma
\end{equation}
using the conservation of comoving entropy. 

We can now introduce the parameter $\neff$, called the \textit{effective number of relativistic species}~\cite{Mangano:2001iu,Bennett:2019ewm,Mangano:2005cc}, which accounts for the non-instantaneous decoupling of neutrinos (in which case we would have $\neff=3$) and the consequent injection of some energy. For example, the energy density of radiation $\rho_r$ is written as the sum of photons and neutrinos:
\begin{equation}
    \rho_r = \left[ 1 + \frac{7}{8} \left( \frac{4}{11} \right)^{\frac{4}{3}} N_{\rm eff} \right] \rho_\gamma
    \label{eq:Neff}
\end{equation}
with $\neff=3.044$~\citep{Mangano:2005cc,2016JCAP...07..051D,Akita:2020szl,Froustey:2020mcq,Bennett:2020zkv,Planck:2018vyg}. Any additional relativistic particle produced before recombination can be treated as an additional contribution to this number (but also other contributions are possible, for example from inflationary gravitational waves~\cite{Giare:2022wxq}). Therefore, observing a $\Delta N_{\rm eff}\neq 0$ could be a hint of new physics. Conversely, smaller values suggest a lower-temperature reheating~\cite{deSalas:2015glj} than expected in the $\lcdm$ Universe. 

Notably, since the radiation energy density $\rho_r$ is proportional to the effective number of neutrinos, different values of $N_{\rm eff}$ modify the sound horizon at recombination. In particular, larger values decrease the horizon and, consequently, require higher values of $H_0$, potentially shifting towards late-time $H_0$ measurements (see, e.g.,~\cite{Vagnozzi:2019ezj}). It has been argued that such a shift would exacerbate the tension with large-scale structure data, but if a proper scale is used to quantify the amplitude of matter fluctuations ($\sigma_{12}$), then the consistency is restored~\cite{Sanchez:2020vvb,Forconi:2025cwp}. 

The strong correlation with the two \textit{pillars} of the Planck vs. ACT inconsistency—the baryon density $\Omega_bh^2$ and the spectral index $n_s$~\cite{Gariazzo:2024sil,Kreisch:2022zxp}—makes $\neff$ an interesting extra parameter to include in this study.

\subsubsection{Early Dark Energy}

Early Dark Energy (EDE) models are a natural hypothesis for dark energy; see, e.g., Refs.~\cite{Wetterich:2004pv,Doran:2006kp,Hollenstein:2009ph,Calabrese:2010uf,Calabrese:2011hg,Calabrese:2011hg,Pettorino:2013ia,Archidiacono:2014msa,Poulin:2023lkg,Poulin:2018dzj,Poulin:2018zxs,Smith:2019ihp,Niedermann:2021vgd,Murgia:2020ryi,Ye:2020btb,Klypin:2020tud,Hill:2020osr,Herold:2021ksg,Herold:2022iib,Reeves:2022aoi,Jiang:2022uyg,Simon:2022adh,Smith:2022hwi,Kamionkowski:2022pkx,Cruz:2023cxy,Cruz:2023lmn,Eskilt:2023nxm,Smith:2023oop,Sharma:2023kzr,Efstathiou:2023fbn,Gsponer:2023wpm,Goldstein:2023gnw}. 
Deviating from the traditional cosmological constant framework, EDE models account for a non-negligible contribution from dark energy in the early Universe. In addition, these EDE models can be based on generic dark energy fluids that are inhomogeneous. Their density and pressure vary over time, leading to a non-static equation of state. 
The phenomenological analyses of these inhomogeneous dark energy models usually require additional dark energy clustering parameters: the dark energy effective sound speed and the dark energy anisotropic stress. The effective sound speed determines the clustering properties of dark energy and consequently affects the growth of matter density fluctuations. Therefore, in principle, its presence could be revealed in large-scale structure observations. The growth of perturbations can also be affected by anisotropic stress contributions, which lead to a damping of velocity perturbations.

Recently, EDE models have garnered significant attention, particularly due to their potential role in addressing some of the aforementioned cosmological tensions~\cite{Kamionkowski:2022pkx,Poulin:2023lkg,Abdalla:2022yfr,Forconi:2023hsj}. Our analysis will concentrate on the EDE implementation detailed in~\cite{Hill:2020osr}. This model proposes that, in the early Universe, a light scalar field deviates from its potential minimum and, constrained by Hubble friction, behaves functionally similar to a cosmological constant. As soon as the Hubble parameter falls below the mass of the field at some particular redshift $z_\star$, the scalar field rolls down its potential and begins to oscillate about the minimum. 
To avoid spoiling late-time cosmology, the vacuum energy must redshift away more quickly than matter (i.e., faster than $a^{-3}$), and the field should remain a subdominant component. A typical set of parameters used in this model includes the fractional contribution to the total energy density of the Universe, $f_{\rm EDE}(z) \equiv \rho_{\rm EDE}(z)/\rho_{\rm tot}(z)$, evaluated at the critical redshift $z_c$ where it reaches its maximum value, and $\theta_i$, which typically describes the initial field displacement. This particular behavior implies a larger amount of energy density in the early Universe (just prior to recombination), a reduction of the sound horizon, and, consequently, a larger value of the Hubble constant inferred from CMB observations. This is why EDE models have been proposed as a possible solution to the Hubble constant tension.

From observations of the primary temperature anisotropies of the CMB with Planck data, there is no evidence for an EDE scenario ($f_{\rm EDE}<0.087$ at $95\%$ CL~\cite{Hill:2020osr}) regardless of the likelihood used~\cite{McDonough:2023qcu}. However, ACT data show a mild preference for nonzero EDE ($f_{\rm EDE}=0.142^{+0.039}_{-0.072}$ at $68\%$ CL~\cite{Hill:2021yec})~\cite{Poulin:2021bjr,Moss:2021obd,Hill:2021yec}.

\section{Statistics} 
\label{sec:method}

Given a set of parameters $\theta$ defining a certain underlying theoretical model $\mscr{M}$, and the observed data $d$, we can apply \textit{Bayes' Theorem}~\cite{D_Agostini_2003,Trotta:2017wnx,Trotta:2008qt}:
\begin{equation}
p(\theta | d) = \frac{p(d|\theta) p(\theta)}{p(d)}\,.
\label{eq:BT}
\end{equation}
In \eq{BT}, $p(\theta | d)$ is the \textit{posterior} probability, which represents our degree of belief about the parameters $\theta$ after observing $d$. The term $p(d|\theta)$, also called the \textit{likelihood} $\mathcal{L}(\theta)$, is the 'probability'\footnote{The likelihood is not a probability distribution over $\theta$, but rather a function of $\theta$ given the observed data~\cite{Trotta:2017wnx}.} of observing the data given the parameters. The factor $p(\theta)$ is called the \textit{prior} and represents all the information we have about the parameters before observing the data; it is common in the literature to choose a flat prior.\footnote{It should be noted that even though we assign a flat prior to a parameter $\theta$, the prior on a function $f(\theta)$ becomes informative.} If this is the case, the posterior becomes functionally identical to the likelihood up to a proportionality constant. On the other hand, any prior knowledge about the parameters is reflected in the posterior.
As a consequence, the best parameters of the model are usually those that maximize the posterior, following the \textit{Maximum a Posteriori} (MAP) method, where $\theta_{MP}$ satisfies $\partial p(\theta|d)/\partial \theta=0$. This contrasts with the \textit{Maximum Likelihood} (ML) method, which finds parameters that maximize the likelihood. For flat, uninformative priors, MAP and ML are identical~\cite{Maltoni:2003cu,Gariazzo:2023joe}. 
Because the posterior is a probability distribution of the parameters, we need a normalization factor, $p(d)$, in \eq{BT}. It is defined as
\begin{equation}
\frac{\int{p(d|\theta)p(\theta)d\theta}}{p(d)}=1\longrightarrow \int{p(d|\theta)p(\theta)d\theta}=p(d)\,.
\label{eq:BEv}
\end{equation}
This term is also called the \textit{Bayesian evidence}. As we will see, it is an important tool for quantifying model performance.

\subsubsection*{Bayesian Evidence}


When comparing models, it is important to take into account the volume of the parameter space and, for example, penalize models with parameters that are unconstrained by the data—an issue that may stem from observational limitations rather than a flaw in the theory. In fact, the 'performance' of a model in fitting the data is highly correlated with its complexity: with a sufficiently high number of parameters, it is possible to accommodate almost any observed data. Therefore, it is crucial to break this correlation by quantifying the necessity for a model not only to fit the data but also to ensure that all parameters are "essential." This principle is encapsulated in the phrase \textit{Pluralitas non est ponenda sine necessitate},\footnote{Attributed to the Franciscan monk William of Ockham (ca. 1285–1349).} or, in other words, the simplest theory compatible with the available evidence ought to be preferred. This is known as the \textit{Occam's razor principle}. 

In statistical terms, a cosmological model $\mscr{M}$ is simply the combination of a set of parameters $\theta$ and their prior distribution, $p(\theta|\mscr{M})$. With this definition in mind, let us reconsider the Bayesian evidence presented in \eq{BEv}~\cite{Trotta:2008qt,Trotta:2017wnx}:
\begin{equation}
p(d|\mscr{M})\equiv\int_{\Omega_\mscr{M}}{p(d|\theta,\mscr{M})p(\theta|\mscr{M})d\theta},
\label{eq:BayesianEv}
\end{equation}
where we have explicitly included the model conditionality. The Bayesian evidence in \eq{BayesianEv} is the average of the likelihood $p(d|\theta,\mscr{M})$ under the prior for a specific model choice. Specifically, it accounts for how well the parameters fit the data while also incorporating information about the change in parameter volume from the prior to the posterior~\cite{Amendola:2015ksp}. Moreover, adding more parameters penalizes the evidence, in accordance with Occam's razor, ensuring that unconstrained parameters do not contribute, as desired.

\subsection{Bayes Factor}
\label{sec:BF}

If we have two different models, $\mscr{M}_0$ and $\mscr{M}_1$, and we want to check whether $\mscr{M}_1$ better describes the data, we can use Bayes' theorem \eq{BT}, assigning a prior to each model, and introduce the \textit{Bayes factor}~\cite{Kass:1995loi,Marshall:2004zd}:
\begin{equation}
    \frac{p(\mscr{M}_0|d)}{p(\mscr{M}_1|d)}=\frac{p(d|\mscr{M}_0)}{p(d|\mscr{M}_1)}\frac{p(\mscr{M}_0)}{p(\mscr{M}_1)}=B_{01}\frac{p(\mscr{M}_0)}{p(\mscr{M}_1)}\,.
    \label{eq:BF}
\end{equation}
In most cases, we do not favor any model, so $p(\mscr{M}_0) = p(\mscr{M}_1)$. If a parameter $\theta_i$ is poorly constrained, then when it varies, the likelihood in \eq{BayesianEv} is approximately constant in the integral. If the prior is factorizable (which is true in almost all cases), the integral decouples and evaluates to unity, meaning it does not contribute to the evidence and consequently to the Bayes factor $B$ in \eq{BF}. This ensures that unconstrained parameters are rightly attributed to poor measurements rather than penalizing the model.

If the Bayes factor $B_{01}$ is greater than one, it suggests that the original model is favored, and we should not switch to $\mscr{M}_1$. On the other hand, if it is smaller than one, we have no evidence to favor $\mscr{M}_0$ over $\mscr{M}_1$. To better quantify the strength of the evidence, we can use the modified~\cite{Trotta:2008qt} Jeffreys' scale~\cite{Jeffreys:1939xee} and compare the Bayes factor with the following thresholds:
\begin{itemize}
    \item $\lvert\ln{B_{01}}\rvert<1$ is ranked as \textit{inconclusive}, and $\mscr{M}_0$ has a posterior probability $<0.750$.
    \item $1\leq\lvert\ln{B_{01}}\rvert<2.5$ is ranked as \textit{weak}, and $\mscr{M}_0$ has a posterior probability of $0.750$.
    \item $2.5\leq\lvert\ln{B_{01}}\rvert<5$ is ranked as \textit{moderate}, and $\mscr{M}_0$ has a posterior probability of $0.923$.
    \item $\lvert\ln{B_{01}}\rvert\geq5$ is ranked as \textit{strong}, and $\mscr{M}_0$ has a posterior probability of $0.993$.
\end{itemize}

If we take two models, $\mscr{M}_0$ and $\mscr{M}_1$, that predict the same value for the parameters $\theta$, then $\mscr{M}_1$ is preferred only if the parameters are strongly constrained by the data due to the integration over the entire prior volume in the Bayesian evidence. In other words, the Bayes factor also measures how the fit performs given the priors (it has a likelihood that is less informative than the prior)~\cite{Trotta:2008qt}.



\subsection{Suspiciousness}

Suppose now that we fix a specific model $\mscr{M}$ and that the data come from two independent datasets, $A$ and $B$.\footnote{New parameters may be introduced when combining independent datasets due to new nuisance parameters or increased constraining power.} If $A$ is described by a set of parameters $\theta_A$ and $B$ by $\theta_B$, then the total evidence can be written as the product of the individual evidences (because the likelihood, but not the posterior, factorizes). Therefore, we can test the confidence in the ability to combine the datasets using the \textit{robustness}~\cite{Amendola:2012wc,Handley:2019wlz,Marshall:2004zd}:
\begin{equation}
    R=\frac{p(A,B)}{p(A)p(B)}
    \label{eq:R}
\end{equation}
where $p(A,B)=\int{p(A|\theta)p(B|\theta)p(\theta)d\theta}$, omitting the conditionality to the model for simplicity. The numerator represents the case where both datasets are explained by the same parameters within the model, while the denominator allows each dataset to be explained by different parameters. The robustness is a ratio of evidences, similar to the Bayes factor \eq{BF}, but applied within a specific model to test the compatibility between datasets. Therefore, the same considerations apply~\cite{Handley:2019wlz}. If $R\gg 1$ ($R\ll 1$), we can interpret it as both datasets being consistent (inconsistent).\footnote{If the datasets are not compatible with one another, since only the likelihoods can be multiplied, we should not expect $R=1$ in this scenario.} 
$R$ is strongly prior-dependent for shared parameters that are constrained; in fact, narrowing the priors decreases the value of $R$. Hence, it must be handled with great care, as an ill-defined prior can indicate an agreement between two datasets that are not actually compatible.\footnote{However, the reverse is not possible due to the unidirectional increase of volume effects.} 
For a better interpretation of $R$~\cite{Handley:2019wlz,Raveri:2018wln,Amendola:2012wc}, we can use the definition of conditional probability and express it as $R=p(A|B)/p(A)$. Thus, we can say that $R$ represents the relative confidence we have in dataset $A$ given knowledge of $B$, compared to the confidence in $A$ alone. From this perspective, it is straightforward that $R>1$ means that $B$ strengthens our confidence in $A$.

If we take the logarithm of \eq{R}, we obtain
\begin{equation}
    \boxed{\log{S}=\log{R}-\log{I}}
\end{equation}
where $S$ is the \textit{Suspiciousness}~\cite{Lemos:2019txn,Handley:2019wlz}, and $I$ is the \textit{Information ratio}. The quantity $I$ accounts for the proportionality of the prior and is defined as $\log{I}=\mcal{D}_{A}+\mcal{D}_{B}-\mcal{D}_{AB}$, where $\mcal{D}$ represents the Kullback-Leibler divergence~\cite{Kullback:1951zyt,Seehars:2014ora,Nicola:2018rcd,Raveri:2016xof,Verde:2013wza}. 
Hence, $S$ is prior-independent~\cite{DES:2020hen} and depends only on the actual mismatch between the posteriors. It is an extremely useful tool for testing the degree of inconsistency between two datasets under different extensions to the standard model, eliminating any possible bias due to prior volume effects~\cite{Handley:2019wlz}.

If the posteriors are such that we may approximate them with a Gaussian (in a broader sense~\cite{Handley:2019wlz,DES:2020hen}), with means and covariance matrices $\mu$ and $\sigma$, then the suspiciousness follows the $\chi^2_d$ distribution, where $d$ is the effective number of degrees of freedom constrained by both datasets~\cite{Handley:2020hdp,Handley:2019wlz,DiValentino:2022rdg,Gariazzo:2023joe}:
\begin{equation}
    \log S = \frac{d}{2}-\frac{\chi^2}{2}
    \label{eq:suspiciousness}
\end{equation}
with 
\begin{equation}
    \chi^2=(\mu_A-\mu_B)^T (\sigma_A+\sigma_B)^{-1} (\mu_A-\mu_B).
    \label{eq:suspiciousness_chi2}
\end{equation}
For an easier interpretation of the result, we can use the inverse cumulative $\chi^2$ distribution to obtain the tension probability:
\begin{equation}
    p=\int^{\infty}_{\chi^2}\frac{x^{d/2-1}e^{-x/2}}{2^{d/2}\Gamma(d/2)}dx\,.
    \label{eq:suspiciousness_p}
\end{equation}
The tension level is then given by
\begin{equation}
    \sigma(p)=\sqrt{2}\,\text{erfc}^{-1}(1-p),
    \label{eq:suspiciousness_sigma}
\end{equation}
which represents the relationship between a given cumulative probability $p$ and its corresponding number of standard deviations $\sigma$ in a normal distribution. Thus, a $3\sigma$ deviation corresponds to $p<0.3\%$, indicating strong tension. The tension probability quantifies the likelihood that the observed tension occurs by chance.

\subsection{Goodness-of-fit}

If we want to understand how \textit{strong} the compromise is when joining two datasets instead of considering them independently, we need to perform the \textit{goodness-of-fit} test~\cite{DES:2020hen,Raveri:2018wln}. This test evaluates the cost of explaining datasets with the same parameter values and is quantified by the estimator:
\begin{multline}
    \qdmap^2=2\ln{\mcal{L}_A(\hat{\theta}_{\rm A})}+2\ln{\mcal{L}_B(\hat{\theta}_{\rm B})}\\
    -2\ln{\mcal{L}_{A+B}(\hat{\theta}_{\rm AB})}
    \label{eq:qdmap}
\end{multline} 
Here, $\hat{\theta}_A$ denotes the parameter values that “best” describe dataset $A$. In frequentist statistics, the parameters are taken such that $\hat{\theta}\equiv\theta_{ML}$, and therefore, $\qdmap\equiv\sqrt{\chi^2}$. On the other hand, in the context of Bayesian analysis, $\hat{\theta}\equiv\theta_{\rm MAP}$. As we have seen, for flat priors, these two cases are identical. 
The test statistic $\qdmap$ is widely used in cosmology (see, e.g.,~\cite{Schoneberg:2021qvd,DES:2020hen,Cruz:2023lmn}). When the likelihoods and posteriors are Gaussian, $\qdmap^2$ follows a $\chi^2$ distribution~\cite{Raveri:2018wln,DES:2020hen}. The goodness-of-fit is expected to degrade by one for each measured parameter and indicates tension if the decrease is significantly higher. Only parameters that are constrained by the data over the prior can contribute to a tension, as prior-constrained parameters cannot be optimized to improve the data fit.



\subsection{AIC}
\label{sec:AIC}

To complete the set of statistical metrics used in this paper, we include the simplest and most widely used Akaike Information Criterion (AIC)~\cite{Akaike} (e.g., in \cite{SolaPeracaula:2017esw,Vagnozzi:2018jhn,Gomez-Valent:2018hwc,Kreisch:2019yzn,Agrawal:2019lmo,Visinelli:2019qqu,RoyChoudhury:2019hls}). The reason for employing this criterion is to facilitate comparisons with results already available in the literature. 

AIC allows for direct and straightforward model comparison, providing a refinement beyond a simple $\chi^2$ difference. As seen from its definition:
\begin{equation}\label{eq:AIC}
    {\rm AIC}=\chi^2+2n_p\,,
\end{equation}
it accounts for the number of fitting parameters ($n_p$). Therefore, it not only evaluates the quality of the fit but also penalizes extra model parameters. We compare the AIC values of non-standard models to that of $\Lambda$CDM by computing the difference $\Delta {\rm AIC}={\rm AIC}_{\Lambda{\rm CDM}}-{\rm AIC}_i$, using the standard model as the benchmark. A positive $\Delta {\rm AIC}$ indicates a preference for the non-standard model. 
More concretely, we quantify this preference using standard thresholds:
\begin{itemize}
    \item If $0 \leq \Delta\textrm{AIC} < 2$, there is \textit{weak evidence} in favor of the new model $i$ compared to the standard model.
    \item If $2 \leq \Delta\textrm{AIC} < 6$, this is considered \textit{positive evidence}.
    \item If $6 \leq \Delta\textrm{AIC} < 10$, we classify it as \textit{strong evidence}.
    \item If $\Delta\textrm{AIC} > 10$, there is \textit{very strong evidence} supporting model $i$ against $\Lambda$CDM.
\end{itemize}

\section{Methods and Dataset}
\label{sec:MD}

\begin{table}
	\begin{center}
		\renewcommand{\arraystretch}{1.5}
		\begin{tabular}{c@{\hspace{0. cm}}@{\hspace{1.5 cm}} c}
			\hline
			\textbf{Parameter}    & \textbf{Prior} \\
			\hline\hline
			$\Omega_{\rm b} h^2$         & $[0.005\,,\,0.1]$ \\
			$\Omega_{\rm c} h^2$     	 & $[0.001\,,\,0.99]$\\
			$100\,\theta_{\rm {MC}}$     & $[0.5\,,\,10]$ \\
			$\tau$                       & $[0.004\,,\,0.8]$\\
			$\log(10^{10}A_{\rm S})$     & $[1.61\,,\,3.91]$ \\
			$n_{\rm s}$                  & $[0.8\,,\, 1.2]$ \\
			\hline
            $\alpha_s$ & $[-1\,,\, 1]$\\
                $w$ & $[-3\,,\, 1]$\\
                $\Omega_k$ & $[-0.4\,,\,0.4]$\\
            $Y_p$ & $[0.1\,,\,0.3]$\\
            $N_{\rm eff}$              & $[1\,,\,5]$\\
            $f_{\rm EDE}$        & $[0\,,\,0.5]$\\
            $\theta_i$               & $[0.1\,,\,3.1]$\\
            $\log_{10}{z_c}$  &$[3.1\,,\,4.3]$\\
            \hline\hline
		\end{tabular}
		\caption{List of the parameter priors used in the MCMC.}
		\label{tab:priors}
	\end{center}
\end{table}

To assess the inconsistency between Planck and ACT, as well as to evaluate the sensitivity of the Hubble tension in different scenarios, we make use of three different datasets:
\begin{itemize}
    \item Planck CMB temperature and polarization power spectra from the legacy Planck release~\cite{Planck:2019nip,Planck:2018vyg} (henceforth Planck). Specifically, we use the combination of the \texttt{Commander} likelihood for temperature TT data at low multipoles ($2\leq\ell\leq 29$), the \texttt{SimAll} likelihood for the same range but with polarization EE, and the \texttt{Plik} high-multipole likelihood for temperature TT, polarization EE, and cross-spectra TE at high multipoles ($30\leq\ell\leq2508$).
    
    \item SH0ES collaboration calibration of the Hubble constant ($H_0=73\pm 1\, \text{km} \text{s}^{-1} \text{Mpc}^{-1}$) using the three-rung distance ladder method with Cepheids~\cite{Riess:2021jrx}.
    
    \item ACT DR4~\cite{ACT:2020frw} + $\tau$ prior ($\mu=0.06$, $\sigma_\tau=0.01$~\cite{ACT:2020gnv}).
\end{itemize}
For all cosmological parameters, we choose flat-prior distributions, varying them uniformly within the conservative ranges listed in \tab{priors}, except for the optical depth, for which we set a Gaussian prior when the dataset in consideration is ACT (as stated above) and when we consider the extension $\Lambda$CDM+BBN, where BBN refers to the BBN prior on $Y_{\rm p}$, as explained in \sect{Yp}. 

For each model, we perform Monte Carlo Markov Chain (MCMC) analyses using the publicly available package \texttt{Cobaya}~\cite{Torrado:2020dgo}, computing the theoretical predictions with \texttt{CAMB}~\cite{Lewis:2007kz,Challinor:2011bk,Howlett:2012mh,Lewis:1999bs} and a modified version of \texttt{CLASS}~\cite{2011arXiv1104.2932L,2011JCAP...07..034B} for EDE~\cite{Hill:2020osr}.\footnote{\href{https://github.com/mwt5345/class_ede?tab=readme-ov-file}{\faGithub\  \texttt{CLASS} EDE}} We explore the posteriors of our parameter space using the MCMC sampler developed for \texttt{CosmoMC}~\cite{Lewis:2002ah}, tailored for parameter spaces with a speed hierarchy, which also implements the “fast dragging” procedure~\cite{neal2005taking}. The convergence of the chains obtained with this procedure is tested using the Gelman-Rubin criterion~\cite{Gelman:1992zz}, with a threshold for chain convergence of $R-1\lesssim 0.02$.

Regarding the computation of the AIC and the goodness-of-fit, we used the minimum $\chi^2$ obtained with the \texttt{py-BOBYQA} minimizer~\cite{2018arXiv180400154C,2018arXiv181211343C,Powell}. In light of the $H_0$ tension, in \eq{qdmap}, we replace dataset B with the SH0ES dataset and assume $\chi^2_{B}=0$~\cite{Schoneberg:2021qvd}. For the Bayes factor, we computed the Bayesian evidence using the publicly available code \texttt{MCEvidence}\footnote{\href{https://github.com/yabebalFantaye/MCEvidence}{\faGithub MCEvidence}}~\cite{Heavens:2017afc}, assigning to $\mathcal{M}_0$ in \eq{BF} the model with the lowest evidence.

To compute the suspiciousness in \eq{suspiciousness}, we assumed uncorrelated datasets. This is not strictly the case for ACT and Planck since, despite being two independent measurements, their multipole ranges overlap. To properly analyze any potential tension between the two, one should follow~\cite{Lemos:2019txn}. However, no joint likelihood exists for these two datasets. Although this approach offers only approximate results, it allows us to assess inconsistencies between the datasets and how they change across different models, without introducing any bias due to prior volume effects. 


\section{Results} 
\label{sec:results}

\begin{table*}[ht]
    \centering
    \renewcommand{\arraystretch}{1.4}
    \setlength{\tabcolsep}{6pt}
    \begin{tabular}{lcccccc}
        \hline
        \hline
        \textbf{Extra-parameter} & $\alpha_s$ &  $w$  & $\Omega_k$ & $Y_\mathrm{P}$  & $N_{\mathrm{eff}}$ & $f_{\mathrm{EDE}}$ \\
        \hline
        \multicolumn{7}{c}{\textbf{Planck}} \\
        \hline
        $n_s$ free  
          & $-0.0060\pm 0.0067$ 
          & $-1.58^{+0.25}_{-0.32}$
          & $-0.045\pm 0.018$   
          & $0.240\pm 0.013$  
          & $2.97^{+0.21}_{-0.19}$  
          & $< 0.0866$ \\
        $n_s=1$    
          & $0.0103\pm0.0067$ 
          & $-1.50^{+0.16}_{-0.21}$
          & $-0.096^{+0.026}_{-0.023}$
          & $0.2929\pm 0.0074$  
          & $3.71\pm 0.11$  
          & $0.135\pm 0.020$ \\
        \hline
        \multicolumn{7}{c}{\textbf{Planck + SH0ES}} \\
        \hline
        $n_s$ free   
          & $-0.0047\pm0.0068$ 
          & $-1.193\pm 0.040$ 
          & $0.0088\pm 0.0020$
          & $0.258\pm 0.013$  
          & $3.48\pm 0.13$  
          & $0.116\pm 0.025$ \\
        $n_s=1$     
          & $0.0092\pm0.0066$
          & $-1.064\pm 0.030$
          & $0.0011\pm 0.0019$
          & $0.2898\pm 0.0073$  
          & $3.71\pm 0.10$  
          & $0.139\pm 0.019$ \\
        \hline
        \multicolumn{7}{c}{\textbf{ACT}} \\
        \hline
        $n_s$ free    
          & $0.060\pm0.028$
          & $-1.16\pm 0.42$ 
          & $-0.003^{+0.020}_{-0.017}$
          & $0.205\pm 0.031$  
          & $2.36\pm 0.43$  
          & $0.148^{+0.057}_{-0.094}$ \\
        $n_s=1$    
          & $0.043\pm0.022$
          & $-1.18\pm 0.43$
          & $0.001^{+0.017}_{-0.015}$
          & $0.226\pm 0.017$  
          & $2.81\pm 0.21$  
          & $0.120^{+0.054}_{-0.078}$ \\
        \hline
        \multicolumn{7}{c}{\textbf{ACT + SH0ES}} \\
        \hline
        $n_s$ free   
          & $0.063\pm0.028$
          & $-1.161\pm 0.063$
          & $0.0102\pm 0.0039$
          & $0.242\pm 0.029$  
          & $3.44\pm 0.23$  
          & $0.123\pm 0.031$ \\
        $n_s=1$  
          & $0.060\pm0.021$
          & $-1.174\pm 0.062$
          & $0.0109\pm 0.0034$
          & $0.220\pm 0.017$  
          & $3.17\pm 0.18$  
          & $0.106\pm 0.027$ \\
        \hline
        \hline
    \end{tabular}
    \caption{Best-fit parameter values at $68\%$ CL for different datasets, with and without imposing an HZ spectrum. The upper bounds are given at $95\%$ CL. These constraints apply to the extra parameter for each alternative model proposed in this paper. Regarding EDE, we report only $f_{\rm EDE}$, as it is better constrained and more indicative of the impact of EDE compared to $\log_{10}z_c$ and $\theta_i$.}
    \label{tab:params}
\end{table*}

\begin{table*}[ht]
    \centering
    \renewcommand{\arraystretch}{1.4}
    \setlength{\tabcolsep}{6pt}
    \begin{tabular}{lcccc}
        \hline
        \hline
        \textbf{Model} 
          & \textbf{Planck} 
          & \textbf{Planck+SH0ES} 
          & \textbf{ACT} 
          & \textbf{ACT+SH0ES} \\
        \hline
        $\Lambda$CDM 
          & $0.9649 \pm 0.0044$
          & $0.9732 \pm 0.0040$
          & $1.008 \pm 0.015$
          & $1.025 \pm 0.015$ \\
        $\Lambda$CDM + $\alpha_s$ 
          & $0.9633 \pm 0.0047$
          & $0.9721 \pm 0.0044$
          & $0.981 \pm 0.020$
          & $0.996 \pm 0.019$ \\
        $w$CDM 
          & $0.9654 \pm 0.0043$
          & $0.9649 \pm 0.0043$
          & $1.007 \pm 0.016$
          & $1.009 \pm 0.016$ \\
        $\Lambda$CDM + $\Omega_k$ 
          & $0.9701 \pm 0.0048$
          & $0.9649 \pm 0.0042$
          & $1.009 \pm 0.017$
          & $1.004 \pm 0.016$ \\
        $\Lambda$CDM + $Y_p$ 
          & $0.9619 \pm 0.0070$
          & $0.9774 \pm 0.0065$
          & $0.977 \pm 0.028$
          & $1.023 \pm 0.026$ \\
        $\Lambda$CDM + BBN 
          & $0.9637 \pm 0.0047$
          & $0.9725 \pm 0.0043$
          & $1.007 \pm 0.016$
          & $1.024 \pm 0.015$ \\
        $\Lambda$CDM + $N_{\mathrm{eff}}$ 
          & $0.9610 \pm 0.0085$
          & $0.9843 \pm 0.0053$
          & $0.961 \pm 0.034$
          & $1.041 \pm 0.017$ \\
        EDE 
          & $0.9693^{+0.0069}_{-0.0082}$
          & $0.9902 \pm 0.0060$
          & $0.994^{+0.039}_{-0.046}$
          & $0.985 \pm 0.032$ \\
        \hline
        \hline
    \end{tabular}
    \caption{Constraints on $n_s$ at $68\%$ CL for the various extensions of the standard model considered in this paper.}
    \label{tab:ns}
\end{table*}

\begin{table*}[ht]
    \centering
    \renewcommand{\arraystretch}{1.4}
    \setlength{\tabcolsep}{6pt}
    \begin{tabular}{lcccc}
        \hline
        \hline
        \textbf{Model} & \textbf{Planck} & \textbf{Planck+SH0ES} & \textbf{ACT} & \textbf{ACT+SH0ES} \\
        \hline
        \multicolumn{5}{c}{$n_s$ free} \\
        \hline
        $\Lambda$CDM
          & $0.02236 \pm 0.00015$
          & $0.02264 \pm 0.00014$
          & $0.02152 \pm 0.00030$
          & $0.02171 \pm 0.00030$ \\[4pt]
        $\Lambda$CDM + $\alpha_s$
          & $0.02240 \pm 0.00016$
          & $0.02267 \pm 0.00015$
          & $0.02137 \pm 0.00032$
          & $0.02156 \pm 0.00032$ \\[4pt]
        $w$CDM
          & $0.02240 \pm 0.00015$
          & $0.02237 \pm 0.00015$
          & $0.02151 \pm 0.00030$
          & $0.02153 \pm 0.00031$ \\[4pt]
        $\Lambda$CDM + $\Omega_k$
          & $0.02258 \pm 0.00017$
          & $0.02236 \pm 0.00015$
          & $0.02149 \pm 0.00032$
          & $0.02151 \pm 0.00032$ \\[4pt]
        $\Lambda$CDM + $Y_p$
          & $0.02229 \pm 0.00021$
          & $0.02273 \pm 0.00019$
          & $0.02111 \pm 0.00042$
          & $0.02169 \pm 0.00041$ \\[4pt]
        $\Lambda$CDM + BBN
          & $0.02233 \pm 0.00015$
          & $0.02262 \pm 0.00014$
          & $0.02150 \pm 0.00031$
          & $0.02169 \pm 0.00030$ \\[4pt]
        $\Lambda$CDM + $N_{\mathrm{eff}}$
          & $0.02227 \pm 0.00022$
          & $0.02283 \pm 0.00015$
          & $0.02098 \pm 0.00046$
          & $0.02185 \pm 0.00032$ \\[4pt]
        EDE
          & $0.02248 \pm 0.00021$
          & $0.02283 \pm 0.00021$
          & $0.02150^{+0.00057}_{-0.00063}$
          & $0.02143 \pm 0.00045$ \\
        \hline
        \multicolumn{5}{c}{$n_s = 1$} \\
        \hline
        $\Lambda$CDM
          & $0.02294 \pm 0.00014$
          & $0.02300 \pm 0.00014$
          & $0.02159 \pm 0.00029$
          & $0.02169 \pm 0.00041$ \\[4pt]
        $\Lambda$CDM + $\alpha_s$
          & $0.02283 \pm 0.00016$
          & $0.02292 \pm 0.00015$
          & $0.02133 \pm 0.00031$
          & $0.02155 \pm 0.00039$ \\[4pt]
        $w$CDM
          & $0.02296 \pm 0.00014$
          & $0.02293 \pm 0.00014$
          & $0.02158 \pm 0.00029$
          & $0.02169 \pm 0.00030$ \\[4pt]
        $\Lambda$CDM + $\Omega_k$
          & $0.02321 \pm 0.00015$
          & $0.02297 \pm 0.00014$
          & $0.02154 \pm 0.00030$
          & $0.02198 \pm 0.00026$ \\[4pt]
        $\Lambda$CDM + $Y_p$
          & $0.02310 \pm 0.00014$
          & $0.02318 \pm 0.00014$
          & $0.02125 \pm 0.00039$
          & $0.02155 \pm 0.00039$ \\[4pt]
        $\Lambda$CDM + BBN
          & $0.02295 \pm 0.00014$
          & $0.02301 \pm 0.00014$
          & $0.02156 \pm 0.00028$
          & $0.02209 \pm 0.00030$ \\[4pt]
        $\Lambda$CDM + $N_{\mathrm{eff}}$
          & $0.02308 \pm 0.00014$
          & $0.02308 \pm 0.00013$
          & $0.02126 \pm 0.00040$
          & $0.02143 \pm 0.00045$ \\[4pt]
        EDE
          & $0.02302 \pm 0.00019$
          & $0.02300 \pm 0.00019$
          & $0.02157 \pm 0.00045$
          & $0.02155 \pm 0.00039$ \\
        \hline
        \hline
    \end{tabular}
    \caption{Constraints on $\Omega_b h^2$ at $68\%$ CL for the various extensions of the standard model considered in this paper.}
    \label{tab:omegab}
\end{table*}

\begin{figure*}[htp]
	\centering
	\includegraphics[width=0.95\textwidth]{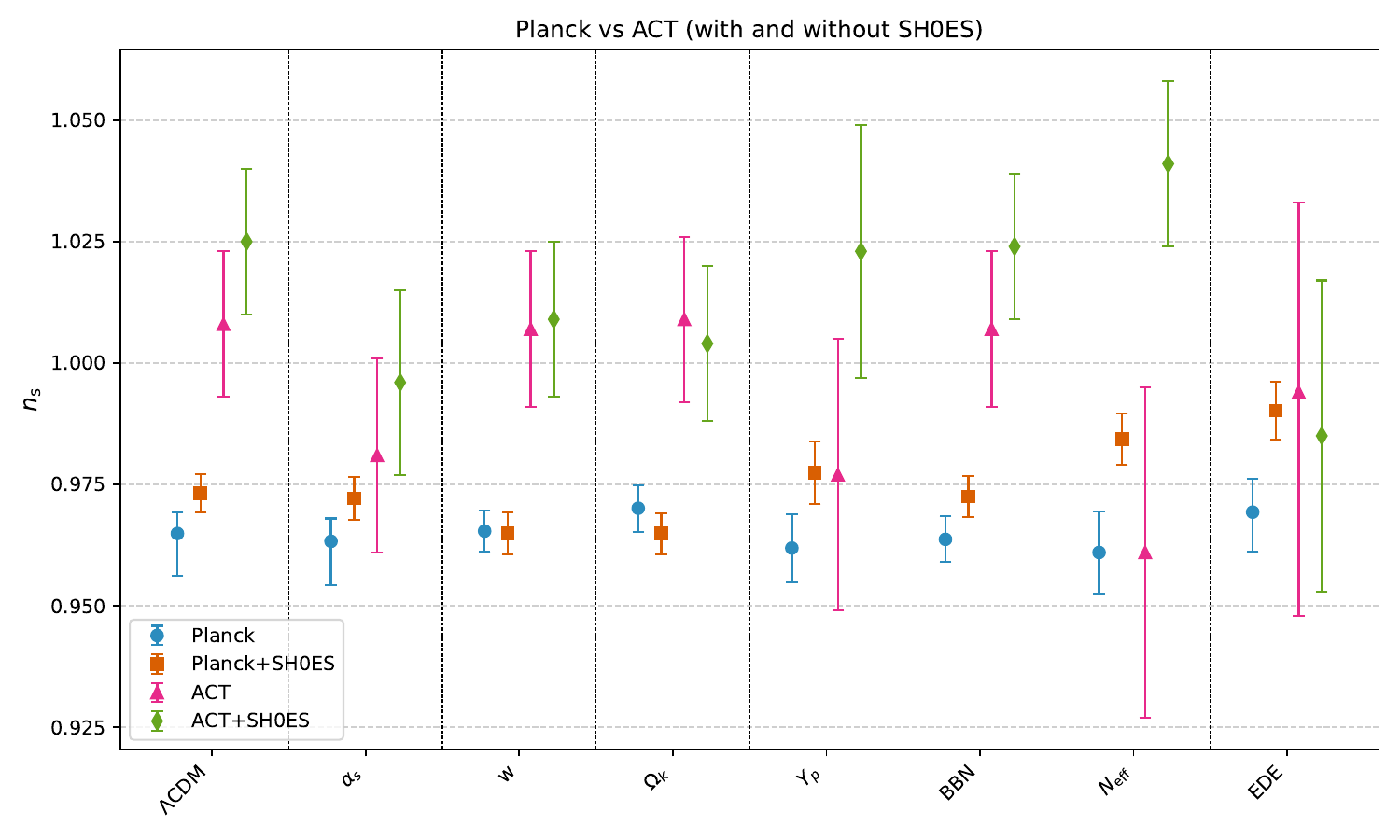}
	\caption{Constraints on $n_s$ at $68\%$ CL for $\Lambda$CDM and its extensions. The constraints are given for Planck and ACT alone, as well as in combination with SH0ES.}
	\label{fig:ns}
\end{figure*}
\begin{figure*}[htp]
	\centering
	\includegraphics[width=0.95\textwidth]{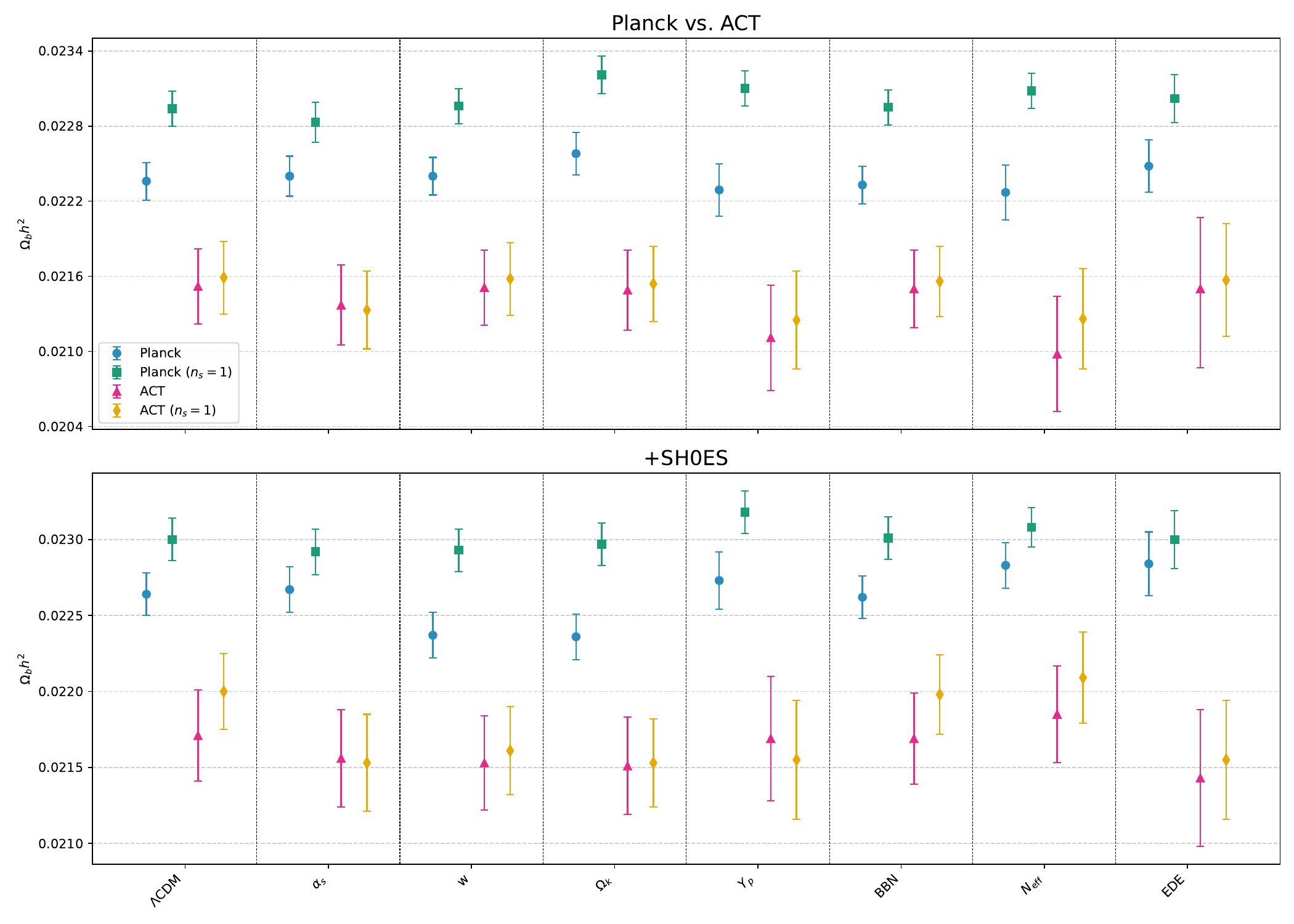}
	\caption{Constraints on $\Omega_b h^2$ at $68\%$ CL for $\Lambda$CDM and its extensions. The upper panel shows the constraints for Planck and ACT alone, while the lower panel presents the results when the data are combined with SH0ES.}
	\label{fig:omb}
\end{figure*}

\begin{figure*}[htp]
	\centering
	\includegraphics[width=0.95\textwidth]{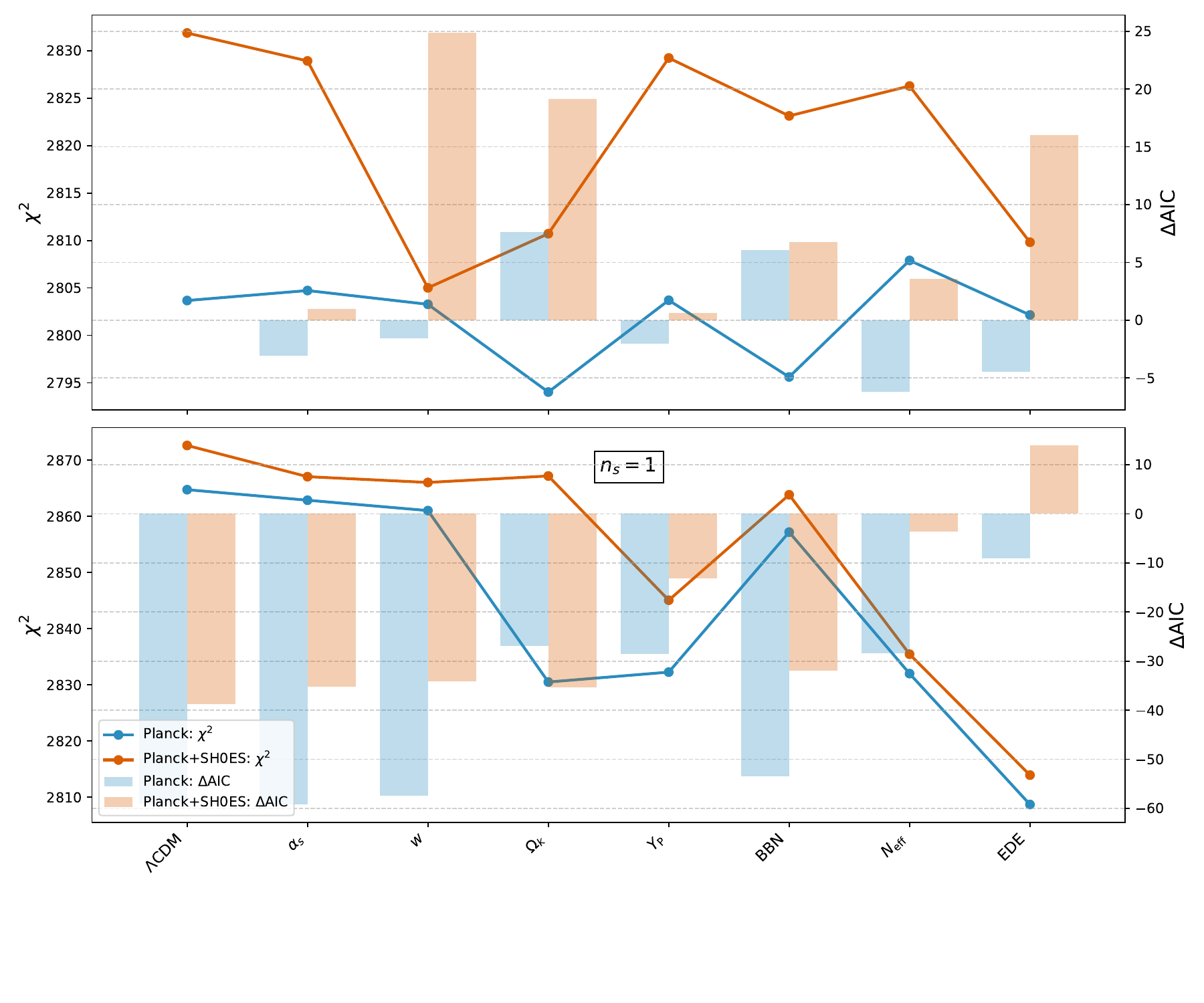}
	\caption{In this graph, the values of $\chi^2$ and $\Delta$AIC are displayed for Planck. The $\chi^2$ values are obtained by minimizing the posterior, as explained in \sect{MD}, while the bars represent the difference in AIC between each model and $\Lambda$CDM, with AIC computed according to \eq{AIC}. Since the Akaike criterion accounts for the number of parameters, it is possible to observe that in some cases, even when the $\chi^2$ of a particular extension is lower than the standard value, $\Delta$AIC remains positive, penalizing the model for the increased number of parameters.}
	\label{fig:AIC_planck}
\end{figure*}
\begin{figure*}[htp]
	\centering
	\includegraphics[width=0.95\textwidth]{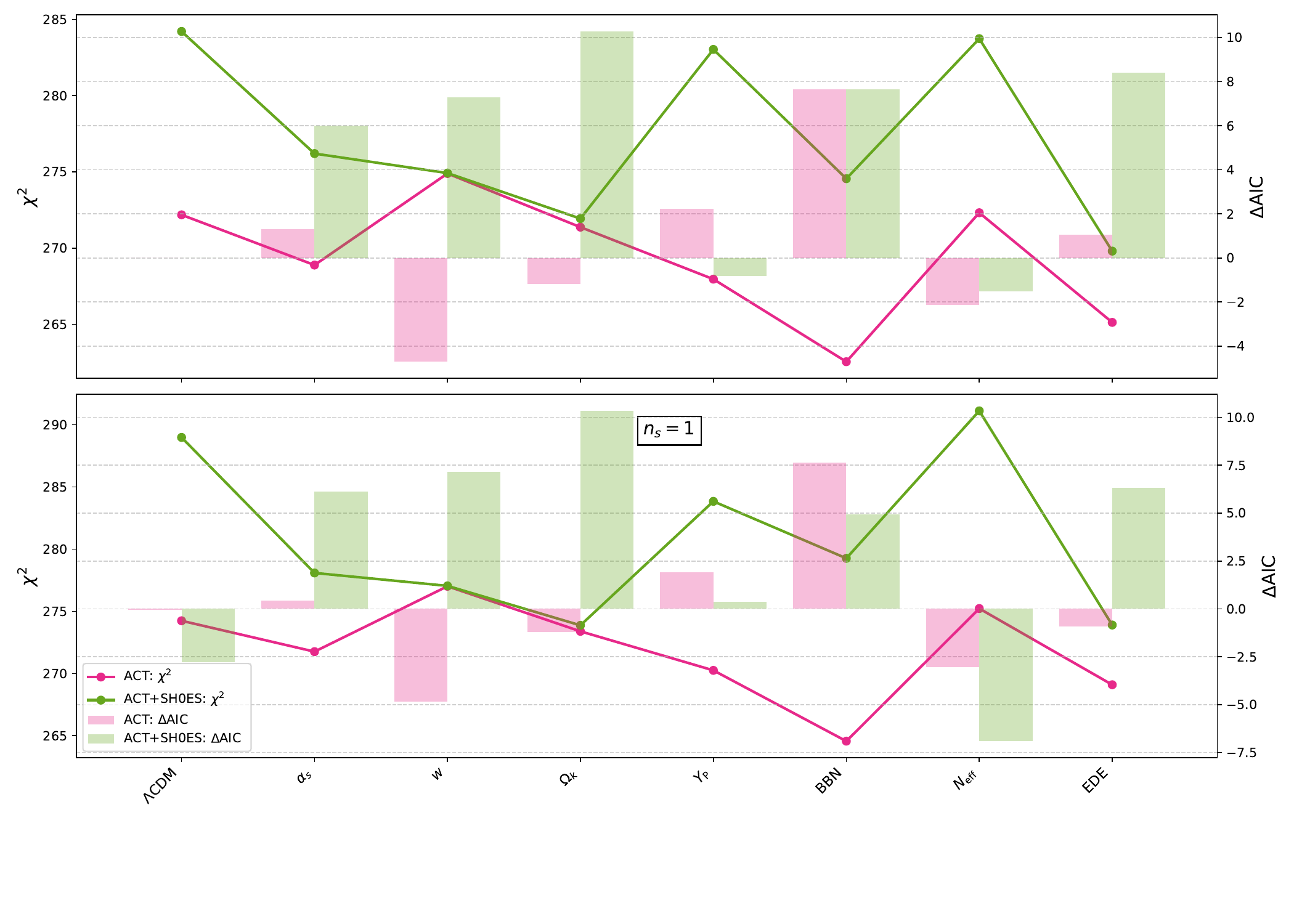}
	\caption{In this graph, the values of $\chi^2$ and $\Delta$AIC are displayed for ACT. The $\chi^2$ values are obtained by minimizing the posterior, as explained in \sect{MD}, while the bars represent the difference in AIC between each model and $\Lambda$CDM, with AIC computed according to \eq{AIC}. Since the Akaike criterion accounts for the number of parameters, it is possible to observe that in some cases, even when the $\chi^2$ of a particular extension is lower than the standard value, $\Delta$AIC remains positive, penalizing the model for the increased number of parameters.}
	\label{fig:AIC_ACT}
\end{figure*}

\begin{table*}[htp]
    \centering
    \renewcommand{\arraystretch}{1.4}
    \setlength{\tabcolsep}{6pt}
    \begin{tabular}{lcccc}
        \hline
        \hline
        \textbf{Model} & \textbf{Planck} & \textbf{Planck+SH0ES} & \textbf{ACT} & \textbf{ACT+SH0ES} \\
        \hline
        \multicolumn{5}{c}{$n_s$ \textbf{free}} \\
        \hline
        $\Lambda$CDM+$\alpha_s$   & -3.05 & 0.94  & 1.30  & 6.01 \\
        $w$CDM                    &  -1.60 & 24.86 &  -4.71  & 7.29 \\
        $\Lambda$CDM+$\Omega_k$   & 7.65 & 19.15 &  -1.19  & 10.28 \\
        $\Lambda$CDM+$Y_p$        &  -2.03 & 0.63  & 2.22  &  -0.82 \\
        $\Lambda$CDM+BBN          & 6.05 & 6.74  & 7.64  & 7.66 \\
        $\Lambda$CDM+$N_{\mathrm{eff}}$ &  -6.22 & 3.59  &  -2.14  &  -1.53 \\
        EDE                       &  -4.48 & 16.05 & 1.06  & 8.41 \\
        \hline
        \multicolumn{5}{c}{$n_s$ \textbf{fixed}} \\
        \hline
        $\Lambda$CDM              & -59.08 & -38.76  &  -0.06  &  -2.77 \\
        $\Lambda$CDM+$\alpha_s$   & -59.21 & -35.20  & 0.42  & 6.13 \\
        $w$CDM                    & -57.36 & -34.18  &  -4.83  & 7.17 \\
        $\Lambda$CDM+$\Omega_k$   & -26.84 & -35.33  &  -1.21  & 10.34 \\
        $\Lambda$CDM+$Y_p$        & -28.60 & -13.21  & 1.93  & 0.37 \\
        $\Lambda$CDM+BBN          & -53.53 & -31.98  & 7.62  & 4.95 \\
        $\Lambda$CDM+$N_{\mathrm{eff}}$ & -28.34 &  -3.60  &  -3.04  &  -6.91 \\
        EDE                       &  -9.03 & 13.93 &  -0.92  & 6.31 \\
        \hline
        \hline
    \end{tabular}
    \caption{$\Delta$AIC values for various models. The AIC, as defined in \eq{AIC}, is compared with the $\Lambda$CDM case. This means that a positive value favors the given model over the standard one. To quantify the \textit{strength} of the preference, we use the standard intervals, as presented in \sect{AIC}.}
    \label{tab:AIC}
\end{table*}

\begin{figure*}[htp]
	\centering
	\includegraphics[width=0.95\textwidth]{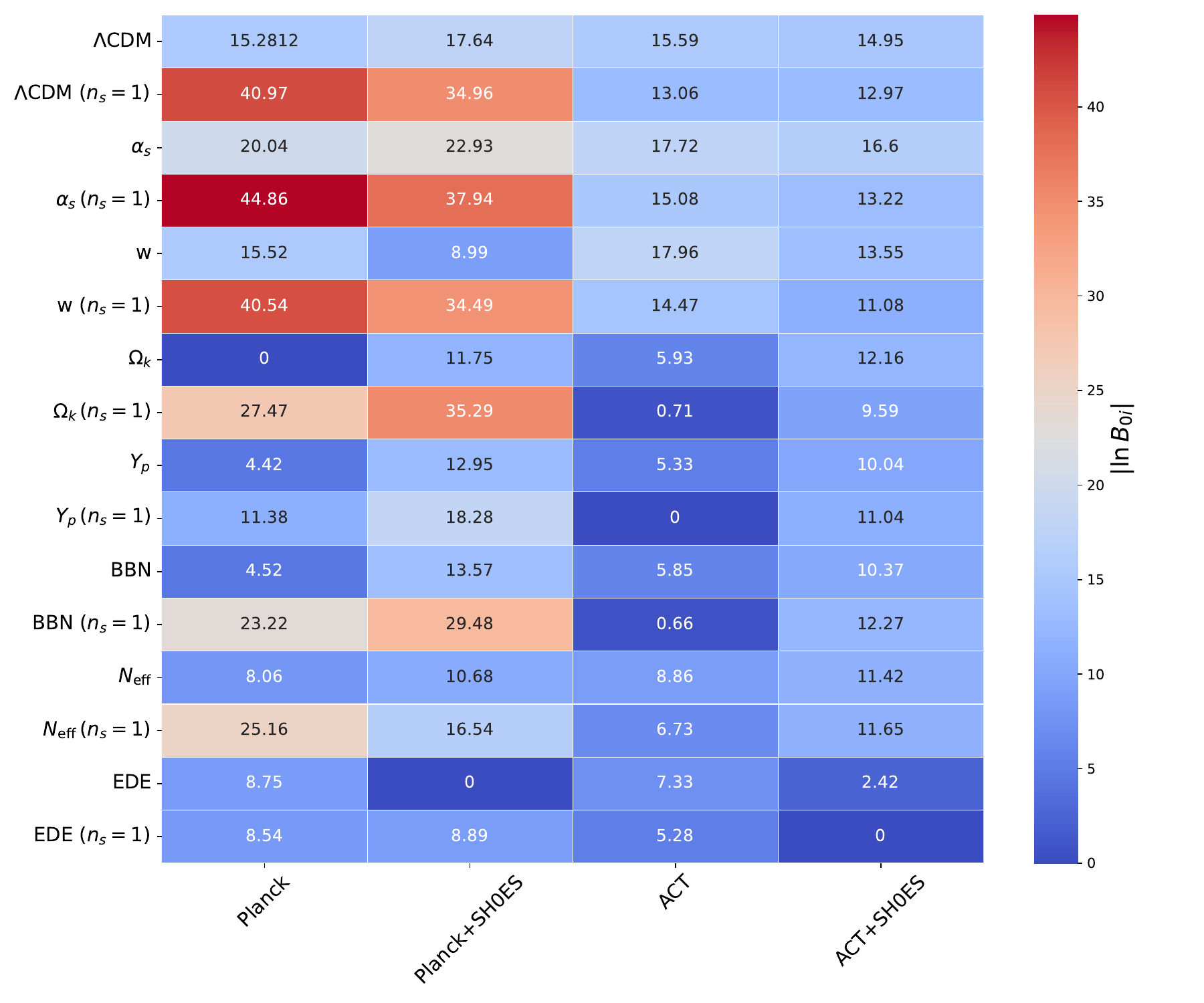}
	\caption{Bayesian factor $B_{01}$ as presented in \eq{BF}. To compute the preference for a model $\mathscr{M}_1$ with respect to $\mathscr{M}_0$, we compare it with the modified Jeffreys' scale (see \sect{BF}). In this case, $\mathscr{M}_0$ is the model with the lowest Bayesian evidence (see \eq{BEv}) and is assigned a Bayesian factor of $0$. Each column represents the Bayesian factor for a specific dataset combination.}
	\label{fig:BF}
\end{figure*}

\begin{figure*}[htp]
	\centering
	\includegraphics[width=0.95\textwidth]{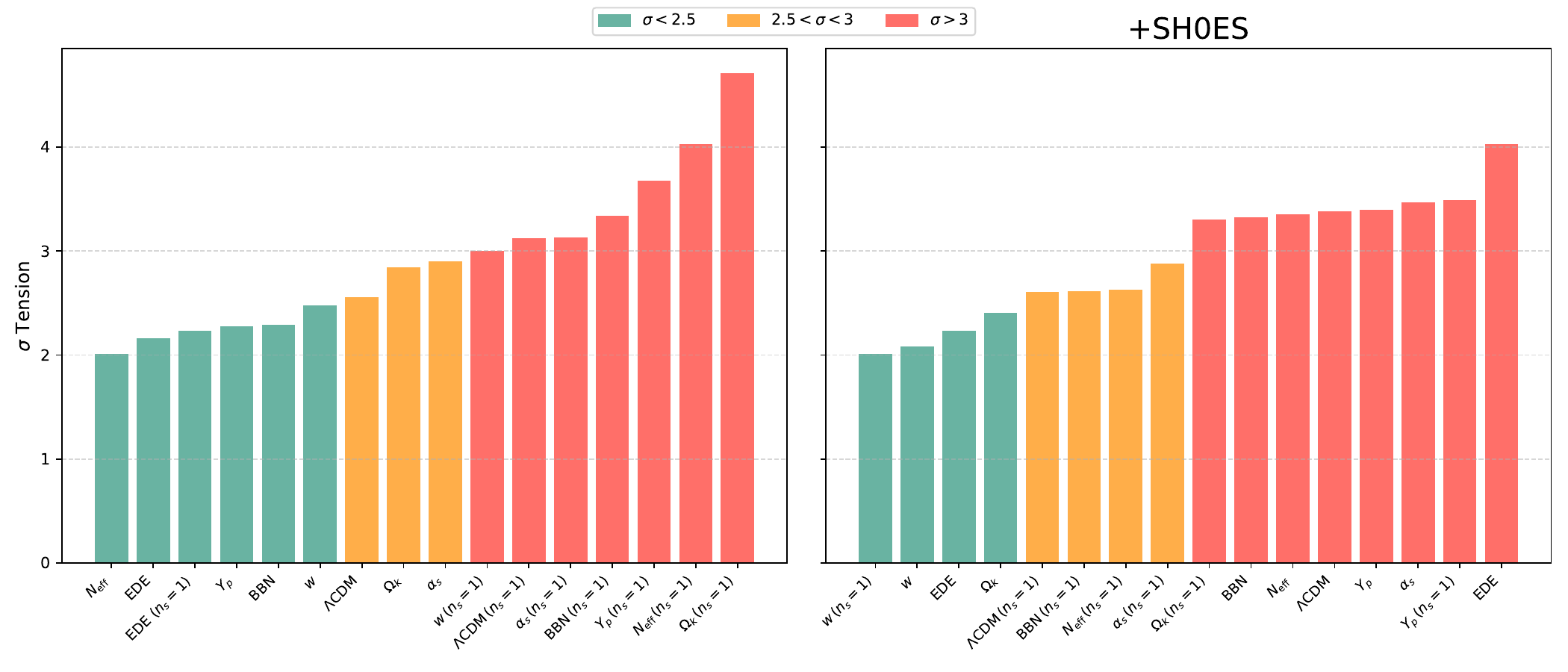}
	\caption{Here, we report the tension level $\sigma(p)$ as computed in \eq{suspiciousness_sigma}, illustrating how the agreement between Planck and ACT varies according to the underlying cosmology.}
	\label{fig:suspiciousness}
\end{figure*}

\begin{table}[htp]
    \centering
    \renewcommand{\arraystretch}{1.4}
    \setlength{\tabcolsep}{6pt} 
    \begin{tabular}{lcccc}
        \hline
        \hline
        \textbf{Model} & $\chi^2$ & logS & p[$\%$] & $\sigma$ \\
        \hline
        \multicolumn{5}{c}{\textbf{Without SH0ES}} \\
        \hline
        \textbf{$n_s$ free} & & & & \\
        $\Lambda$CDM & 16.67 & -5.33 & 1.06 & 2.56 \\
        $\Lambda$CDM+$\alpha_s$ & 21.02 & -7.01 & 0.37 & 2.90 \\
        $w$CDM & 17.75 & -5.38 & 1.31 & 2.48 \\
        $\Lambda$CDM+$\Omega_k$ & 20.55 & -6.77 & 0.45 & 2.84 \\
        $\Lambda$CDM+$Y_p$ & 16.24 & -4.62 & 2.30 & 2.27 \\
        $\Lambda$CDM+BBN & 16.34 & -4.67 & 2.22 & 2.29 \\
        $\Lambda$CDM+$N_{\rm eff}$ & 14.42 & -3.71 & 4.42 & 2.01 \\
        EDE & 18.41 & -4.71 & 3.07 & 2.16 \\
        \hline
        \textbf{$n_s$ fixed} & & & & \\
        $\Lambda$CDM & 19.19 & -7.09 & 0.18 & 3.13 \\
        $\Lambda$CDM+$\alpha_s$ & 21.11 & -7.56 & 0.17 & 3.13 \\
        $w$CDM & 20.10 & -7.05 & 0.27 & 3.00 \\
        $\Lambda$CDM+$\Omega_k$ & 36.25 & -15.12 & 0.0002 & 4.71 \\
        $\Lambda$CDM+$Y_p$ & 25.88 & -9.94 & 0.023 & 3.68 \\
        $\Lambda$CDM+BBN & 22.88 & -8.44 & 0.084 & 3.34 \\
        $\Lambda$CDM+$N_{\rm eff}$ & 29.17 & -11.58 & 0.0061 & 4.03 \\
        EDE & 17.48 & -4.74 & 2.54 & 2.23 \\
        \hline
        \multicolumn{5}{c}{\textbf{With SH0ES}} \\
        \hline
        \textbf{$n_s$ free} & & & & \\
        $\Lambda$CDM & 23.25 & -8.63 & 0.072 & 3.38 \\
        $\Lambda$CDM+$\alpha_s$ & 25.89 & -9.44 & 0.053 & 3.47 \\
        $w$CDM & 14.90 & -3.95 & 3.72 & 2.08 \\
        $\Lambda$CDM+$\Omega_k$ & 17.22 & -5.11 & 1.60 & 2.41 \\
        $\Lambda$CDM+$Y_p$ & 25.22 & -9.12 & 0.069 & 3.39 \\
        $\Lambda$CDM+BBN & 24.64 & -8.82 & 0.088 & 3.33 \\
        $\Lambda$CDM+$N_{\rm eff}$ & 24.87 & -8.94 & 0.080 & 3.35 \\
        EDE & 18.94 & -4.97 & 2.57 & 2.23 \\
        \hline
        \textbf{$n_s$ fixed} & & & & \\
        $\Lambda$CDM & 15.30 & -5.15 & 0.92 & 2.61 \\
        $\Lambda$CDM+$\alpha_s$ & 19.13 & -6.56 & 0.40 & 2.88 \\
        $w$CDM & 14.42 & -3.71 & 4.42 & 2.01 \\
        $\Lambda$CDM+$\Omega_k$ & 22.59 & -8.29 & 0.095 & 3.31 \\
        $\Lambda$CDM+$Y_p$ & 24.20 & -9.10 & 0.048 & 3.49 \\
        $\Lambda$CDM+BBN & 17.09 & -5.54 & 0.90 & 2.61 \\
        $\Lambda$CDM+$N_{\rm eff}$ & 17.17 & -5.58 & 0.87 & 2.62 \\
        EDE & 33.22 & -12.61 & 0.0056 & 4.03 \\
        \hline
        \hline
    \end{tabular}
    \caption{The compatibility between Planck and ACT has been computed using the suspiciousness $\log S$ defined in \eq{suspiciousness}. To better estimate the tension, we also report the $\chi^2$ value as defined in \eq{suspiciousness_chi2}, the probability $p$ (see \eq{suspiciousness_p}), and the tension level $\sigma(p)$ (as in \eq{suspiciousness_sigma}).}
    \label{tab:suspiciousness}
\end{table}

\begin{figure*}[htp]
	\centering
	\includegraphics[width=0.6\textwidth]{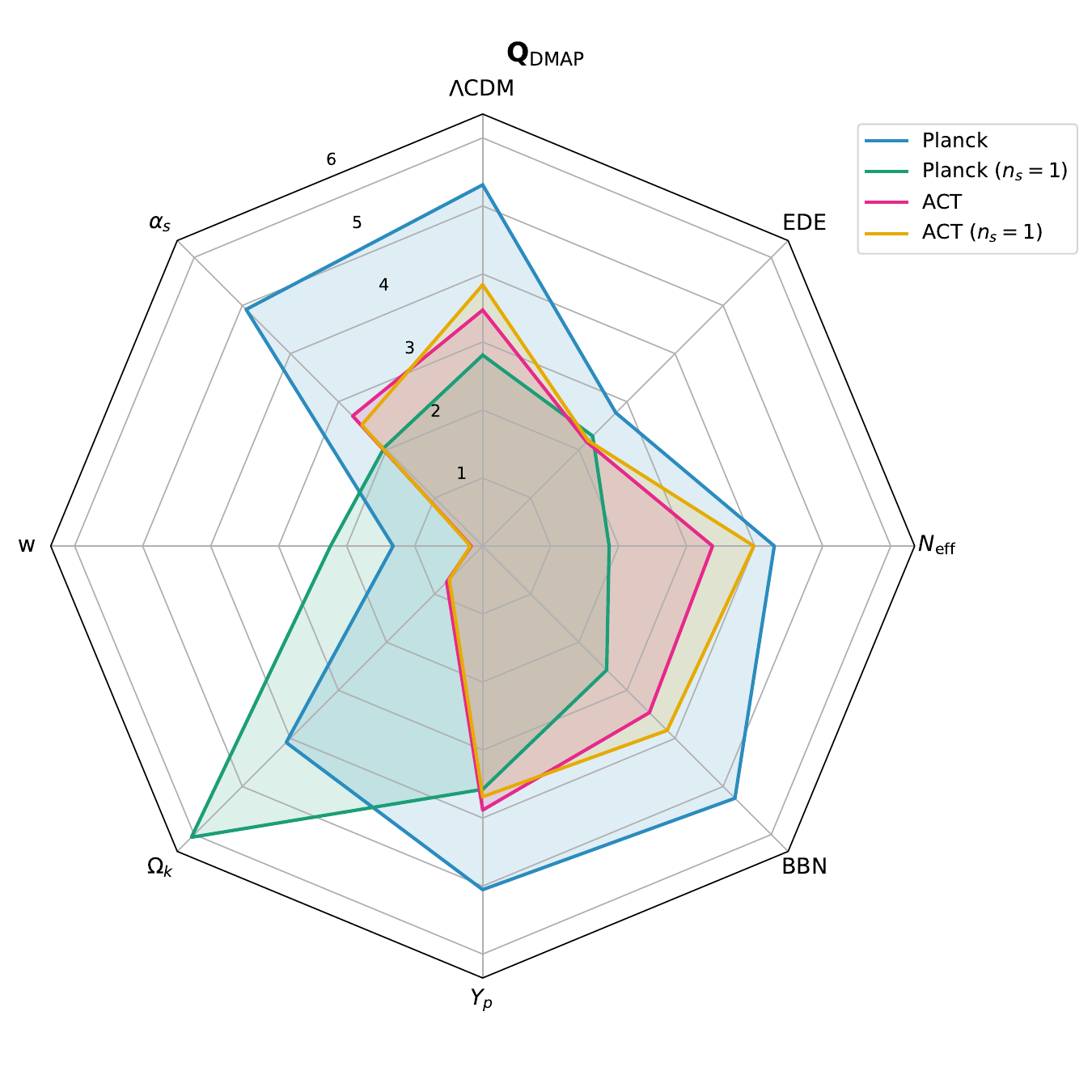}
	\caption{With this plot, we represent the $\qdmap$ values for the different models. $\qdmap$ has been computed following \eq{qdmap} and assigning $\chi^2_{\rm SH0ES} \equiv 0$, as explained in \sect{MD}.}
	\label{fig:qdmap}
\end{figure*}
\begin{table}[ht]
    \centering
    \renewcommand{\arraystretch}{1.4}
    \setlength{\tabcolsep}{6pt} 
    \begin{tabular}{lcc}
        \hline
        \hline
        \textbf{Model} & \textbf{Planck} & \textbf{ACT} \\
        \hline
        \multicolumn{3}{c}{\textbf{$n_s$ free}} \\
        \hline
        $\Lambda$CDM                & 5.31 & 3.47 \\
        $\Lambda$CDM + $\alpha_s$   & 4.92 & 2.70 \\
        $w$CDM                      & 1.32 & 0.17 \\
        $\Lambda$CDM + $\Omega_k$     & 4.09 & 0.75 \\
        $\Lambda$CDM + $Y_p$          & 5.05 & 3.88 \\
        $\Lambda$CDM + BBN          & 5.24 & 3.47 \\
        $\Lambda$CDM + $N_{\mathrm{eff}}$ & 4.29 & 3.38 \\
        EDE                         & 2.77 & 2.16 \\
        \hline
        \multicolumn{3}{c}{\textbf{$n_s$ fixed}} \\
        \hline
        $\Lambda$CDM                & 2.81 & 3.84 \\
        $\Lambda$CDM + $\alpha_s$   & 2.05 & 2.52 \\
        $w$CDM                      & 2.24 & 0.18 \\
        $\Lambda$CDM + $\Omega_k$     & 6.06 & 0.69 \\
        $\Lambda$CDM + $Y_p$          & 3.58 & 3.69 \\
        $\Lambda$CDM + $BBN$          & 2.58 & 3.83 \\
        $\Lambda$CDM + $N_{\mathrm{eff}}$ & 1.86 & 3.99 \\
        EDE                         & 2.29 & 2.19 \\
        \hline
        \hline
    \end{tabular}
    \caption{Here, we report the $\qdmap$ values for the different models, with and without fixing an HZ spectrum. $\qdmap$ measures the compatibility of explaining the data by combining two datasets instead of considering them independently. In this case, the comparison has been performed between Planck/ACT and SH0ES, thus quantifying the Hubble tension. $\qdmap$ has been computed using \eq{qdmap} and assigning $\chi^2_{\rm SH0ES} \equiv 0$, as explained in \sect{MD}.}
    \label{tab:qdmap}
\end{table}

In \tab{params}, the constraints on the parameters beyond the standard model are reported. In \tab{ns} and \fig{ns}, the estimates for the spectral index $n_s$ can be found, whereas in \tab{omegab} and \fig{omb}, those for $\Omega_b h^2$ are presented. To analyze the performance of a model with respect to the others, given a dataset combination, we can study the $\chi^2$ and $\Delta$AIC values in \fig{AIC_planck}, \fig{AIC_ACT}, and \tab{AIC}. For further comparison, the Bayes factors are reported in \fig{BF}. To assess the Planck and ACT discrepancy, the tension levels are depicted in \fig{suspiciousness}, and all Suspiciousness parameters are reported in \tab{suspiciousness}. Lastly, to evaluate the sensitivity of the Hubble tension to the underlying cosmology, we examine the $\qdmap$ values in \tab{qdmap} and \fig{qdmap}.

Before analyzing the performance of each model separately, we first present a general overview of model performance according to the Bayes factor defined in \eq{BF}. As anticipated in \sect{MD}, zero values correspond to the cases with the lowest Bayesian evidence ($B_{00}=0$). 
For Planck-only data, $\Lambda$CDM+$\Omega_k$ is the model favored by the data. Only the Helium-fraction alternative (both with and without the BBN prior) yields a Bayes factor indicative of a \textit{moderate} preference for $\mathscr{M}_0$ over $\mathscr{M}_1$. In every other extension, the evidence strongly supports $\Omega_k$, with a posterior probability exceeding $99.3\%$. 
If we include SH0ES data, the non-flat geometry is strongly disfavored, as are all the other alternatives, in favor of the EDE scenario. This result is expected, as EDE was introduced to address the Hubble tension. Notably, imposing an HZ spectrum generally worsens the Bayesian factor for most models, although EDE with Planck alone experiences a slight improvement.

In the case of ACT-only data, we observe smaller overall Bayes factors but still see no clear \textit{weak} or \textit{moderate} preferences, except in a few instances. Imposing $n_s=1$ consistently reduces the evidence for most extensions, with the following exceptions:
\begin{align*}
    &|\ln{B_{01}}|=10.04\xrightarrow{n_s=1}11.04\quad\text{($\Lambda$CDM+$Y_p$)}\\
    &|\ln{B_{01}}|=10.37\xrightarrow{n_s=1}12.27\quad\text{($\Lambda$CDM+BBN)}\\
    &|\ln{B_{01}}|=11.42\xrightarrow{n_s=1}11.65\quad\text{($\Lambda$CDM+$N_{\rm eff}$)}
\end{align*}
for ACT+SH0ES. ACT-only data are insufficient to decisively distinguish among the extensions with $Y_p$, $Y_p$+BBN prior, and $\Omega_k$, all combined with $n_s=1$. Again, the inclusion of SH0ES amplifies the preference for EDE, as EDE+$n_s=1$ is now the preferred model according to the data, with only a \textit{weak} preference against EDE alone, while all other alternatives remain strongly disfavored. 

If we now look at \fig{ns}, it is immediately apparent that Planck constraints on $n_s$ remain unaffected, but the set of $\{\alpha_s,Y_p,N_{\rm eff},\text{EDE}\}$ alternatives shifts ACT values and increases error bars, allowing an observational consistency below $2\sigma$. Specifically, we have $n_s=0.981\pm0.020$, $n_s=0.977\pm0.028$, $n_s=0.961\pm0.034$ and $n_s=0.994^{+0.039}_{-0.046}$ at 68\% CL for each respective model. 
The inclusion of SH0ES increases the expectation values for both Planck and ACT and enhances their overlap, with the exception of EDE. On the other hand, the baryon parameter is less sensitive to the different models, and an agreement is almost never recovered. The HZ spectrum does increase the $\Omega_b h^2$ value for both ACT and Planck; however, the excursion for ACT is not comparable to that of Planck, thus exacerbating this parameter tension.

\subsubsection{$\Lambda$CDM}

From previous work (see, e.g.,~\cite{Handley:2020hdp,DiValentino:2022rdg, Hazra:2024nav}), it is known that within the standard cosmological model, there is a $\sim2.6\sigma$ tension level (see \eq{suspiciousness_sigma}) between Planck and ACT measurements. This level serves as a benchmark for all other cosmological scenarios considered here: if the Planck–ACT agreement for a given extension lies below this threshold, the model appears promising for alleviating the tension. Adding the SH0ES prior exacerbates the tension to more than $3\sigma$, as seen in \fig{suspiciousness}. If we reduce the number of free parameters, i.e., fix $n_s$ to unity, the results are inverted. In fact,
\begin{align*}
    &\log{S} = -5.33 \xrightarrow{n_s=1}  -7.09\quad \text{(Planck)} \\
    &\log{S} = -8.63 \xrightarrow{n_s=1}  -5.15\quad \text{(Planck+SH0ES)} 
\end{align*}

Since $\Lambda$CDM is, by definition, not suitable for solving the Hubble tension, we adopt its $\qdmap$ value as our reference: $\qdmap^\Lambda=5.31$ for Planck-only analyses. This value can decrease when additional datasets are incorporated (see, e.g.,~\cite{Schoneberg:2021qvd}). Replacing Planck with ACT data yields a smaller tension, with $\qdmap=3.47$, signifying a comparatively moderate discrepancy. Of particular interest is the role of the HZ spectrum: it substantially lowers the discrepancy for Planck, yet slightly increases it for ACT. Specifically,
\begin{align*}
   & \qdmap = 5.31 \xrightarrow{n_s=1}2.81\quad \text{(Planck)}\\
   & \qdmap = 3.47 \xrightarrow{n_s=1}3.84\quad \text{(ACT)}
\end{align*}
Imposing an HZ spectrum appears to mitigate the tension in Planck, while no similar behavior is observed in ACT. Results are shown in \fig{qdmap} and reported in \tab{qdmap}. 

However, even though fixing $n_s$ is a promising approach for reconciling Planck and SH0ES, it is strongly disfavored by $\Delta$AIC, while no significant penalty is given when using ACT-only data. However, once SH0ES is included, the Akaike criterion favors the standard model over the HZ counterpart. The variation in AIC (see \eq{AIC}) is:
\begin{align*}
    &\Delta\text{AIC}=-59.08\xrightarrow{+SH0ES}-38.76\quad\text{(Planck)} \\
    & \Delta\text{AIC}=-0.06\xrightarrow{+SH0ES}-2.77\quad\text{(ACT)}
\end{align*}

\subsubsection{$\Lambda$CDM$+\alpha_s$}

Introducing a scale dependency to the spectral index does not alleviate the Planck-ACT tension. In fact, for all four combinations (with or without SH0ES, and with or without a fixed $n_s$), the suspiciousness $\log S$ decreases, reaching its most negative value, $\log S= -9.44$, when $n_s$ is free and SH0ES is included; this corresponds to a tension probability of $0.053\%$. 

Although a running spectral index can slightly reduce the tension with SH0ES relative to $\Lambda$CDM, it is not the best model to tackle this tension. However, imposing $n_s=1$ promotes this scenario to the second-best solution to the Hubble tension for Planck and also yields a moderate reduction in tension for ACT. In particular, we have
\begin{align*}
    & \qdmap = 4.92\xrightarrow{n_s=1}2.05\quad\text{(Planck)}\\
    & \qdmap = 2.70\xrightarrow{n_s=1}2.52\quad\text{(ACT)}
\end{align*}

Regarding AIC, $\alpha_s$ is only weakly preferred when fitting ACT data alone ($\Delta$AIC = 1.30) or Planck+SH0ES ($\Delta$AIC = 0.94), but becomes strongly favored once SH0ES is included ($\Delta$AIC = 6.01) and disfavored for Planck-only data ($\Delta$AIC = -3.05). When $n_s$ is held fixed, the model’s performance is nearly identical to $\Lambda$CDM for both Planck cases, as seen in \tab{AIC}. In contrast, ACT+SH0ES yields strong support for running, similar to the free-$n_s$ case.

By looking at \tab{params}, we notice that across all datasets, the parameter $\alpha_s$ remains consistent with zero at $1\sigma$ for Planck, regardless of whether $n_s$ is allowed to vary. For the remaining data combinations, $\alpha_s=0$ is compatible within $2\sigma$.

\subsubsection{$w$CDM}

Replacing the cosmological constant with a dark energy equation of state $w$, we increase the number of parameters from $6$ to $7$. Therefore, despite $\chi^2$ (see \eq{suspiciousness_chi2}) between the two datasets being larger compared to the standard model ($\chi^2_w=17.75$ instead of $\chi^2_\Lambda=16.67$), the tension probability is slightly higher: $p=1.3\%$. When $n_s$ is fixed, $w$CDM performs poorly at resolving the discrepancy between Planck and ACT, but once SH0ES data are included, the situation changes:
\begin{align*}
    &\log{S}=-5.38\xrightarrow{+SH0ES}-3.95\quad (n_s\,\text{free})\\
    &\log{S}=-7.05\xrightarrow{+SH0ES}-3.71\quad (n_s\,\text{fixed})
\end{align*}
Both with and without an HZ spectrum, the tension levels remain at $\sim 2\sigma$, as seen in \fig{suspiciousness}.

This dark energy alternative offers one of the best scenarios for reconciling CMB and SH0ES measurements. The $\qdmap$ values are the lowest for Planck ($\qdmap=1.32$), ACT ($\qdmap=0.17$), and ACT with $n_s=1$ ($\qdmap=0.18$). The only exception is Planck with a fixed spectral index, which gives the third-lowest value: $2.24$. 

This model performs best when SH0ES is included, showing very strong ($\Delta^{\rm Planck}$AIC$=24.86$) and strong ($\Delta^{\rm ACT}$AIC$=7.29$) evidence against $\Lambda$CDM. However, for Planck-only or ACT-only analyses, $\Lambda$CDM remains weakly or positively preferred, respectively. Introducing an HZ spectrum preserves the viability of $w$CDM for ACT and ACT+SH0ES but leads to more negative values for Planck-based datasets:
\begin{align*}
    &\Delta\text{AIC}=-1.60\xrightarrow{n_s=1}-57.36\quad\text{(Planck)}\\
    &\Delta\text{AIC}=24.86\xrightarrow{n_s=1}-34.18\quad\text{(Planck+SH0ES)}
\end{align*}

For ACT alone with a free spectral index, $w=-1.16\pm 0.42$ at $68\%$ CL, and with $n_s=1$, $w=-1.18\pm 0.43$ at $68\%$ CL; both results are consistent with a cosmological constant, which is recovered when $w=-1$. Meanwhile, Planck+SH0ES excludes $w=-1$ at over $3\sigma$ ($w=-1.193\pm 0.040$), though this significance diminishes to slightly above $2\sigma$ ($w=-1.064\pm 0.030$) under the fixed-$n_s$ assumption.

\subsubsection{$\Lambda$CDM$+\Omega_k$}

Although a nonzero curvature appears to be favored by Planck alone (yielding the lowest Bayesian evidence), it does not reconcile the Planck and ACT datasets. This is particularly evident under the HZ assumption, with
\begin{align*}
    & \log{S}=-15.12,\,\sigma(p)=4.71\quad\text{(without SH0ES)}\\
    & \log{S}=-8.29,\,\sigma(p)=3.31\quad\text{(with SH0ES)}
\end{align*}
Conversely, if $n_s$ is an extra parameter of the theory, the tension probability more than doubles relative to a flat universe model, resulting in the second-lowest $\chi^2$ among models with the same number of parameters, and a suspiciousness of $\log S= -5.11$ once SH0ES is included (though it decreases notably if SH0ES is excluded, see \tab{suspiciousness}).

For Planck, allowing curved geometry slightly reduces the tension with SH0ES ($\qdmap=4.09$), whereas fixing $n_s=1$ yields the largest discrepancy ($\qdmap>6$). Conversely, for ACT, relaxing flatness lowers the tension substantially (below unity) in both scenarios.

According to the AIC statistics, introducing $\Omega_k$ is strongly preferred by Planck and very strongly preferred by Planck+SH0ES and ACT+SH0ES. Specifically,
\begin{align*}
    &\Delta\text{AIC}=7.65\xrightarrow{+SH0ES}19.15\quad\text{(Planck)}\\
    &\Delta\text{AIC}=-1.19\xrightarrow{+SH0ES}10.28\quad\text{(ACT)}\,.
\end{align*}
According to the results in \tab{AIC}, \fig{AIC_planck}, and \fig{AIC_ACT}, this preference only persists when $n_s=1$ in the ACT+SH0ES case; elsewhere, it is disfavored. 

Planck's constraints on curvature~\cite{Planck:2018vyg} suggest a slight preference for a closed geometry, though consistent with flat space at $\sim 2\sigma$. In contrast, ACT measurements provide tighter bounds on curvature~\cite{ACT:2020gnv}, showing perfect compatibility with $\Omega_k=0$. Including SH0ES data tightens both Planck and ACT posteriors, excluding perfect flatness at over $4\sigma$ for Planck and $2\sigma$ for ACT. Fixing $n_s$ does not significantly affect the results for ACT. However, Planck alone shifts to $\Omega_k=-0.096^{+0.026}_{-0.023}$ at $68\%$ CL, further increasing the discrepancy with flatness. On the other hand, Planck+SH0ES is now compatible with $\Omega_k=0$, yielding $\Omega_k= 0.0011\pm0.0019$, as reported in \tab{params}.

\subsubsection{$\Lambda$CDM$+Y_{\mathrm{P}}$}

Extending $\Lambda$CDM by allowing $Y_p$ to vary can yield the second-highest suspiciousness, $\log{S}\sim4.6$, in the simple case of free $n_s$ and no SH0ES, largely independent of whether a BBN prior is applied. The same independence to the Helium prior is maintained when SH0ES is included, but it leads to a disfavored tension level ($\sim 3.3\sigma$), similar to the standard $\Lambda$CDM one. However, the other two combinations with an HZ spectrum indicate a preference for including the BBN prior:
\begin{align*}
    & \log{S}=-9.94\xrightarrow{+\rm BBN\,prior}-8.44\quad \text{(without SH0ES)}\\
    & \log{S}=-9.10\xrightarrow{+\rm BBN\,prior}-5.54\quad \text{(with SH0ES)}
\end{align*}

In terms of $\qdmap$, Planck yields $\qdmap=5.05$ and ACT $\qdmap=3.88$, comparable to the base results. Fixing $n_s$ does not alleviate the tension for Planck, as $\qdmap$ diminishes less effectively than in the standard scenario; for ACT, it remains near the $\Lambda$CDM value:
\begin{align*}
    & \qdmap = 5.05\xrightarrow{n_s=1}3.58\quad\text{(Planck)}\\
    & \qdmap = 3.88\xrightarrow{n_s=1}3.69\quad\text{(ACT)}\,.
\end{align*}
Imposing a BBN prior likewise does not substantially shift $\qdmap$, except that it restores a more standard behavior for Planck with $n_s=1$.

$\Lambda$CDM+$Y_{\mathrm{P}}$ generally tracks $\Lambda$CDM in terms of the $\Delta$AIC results reported in \tab{AIC}. Introducing a BBN prior, however, yields strong evidence for all four data combinations if $n_s$ is free. Instead, if we fix $n_s$, only ACT data favor this extension, with positive values such as $\Delta$AIC$=4.95$ (with SH0ES) and $\Delta$AIC$=7.62$ (without SH0ES).

At $68\%$ CL, Planck constrains $Y_{\mathrm{P}} = 0.240\pm0.013$, which increases to $Y_{\mathrm{P}} = 0.2929\pm0.0074$ when $n_s=1$; notably, this is higher than typical BBN predictions. The increase is less pronounced with ACT, where $Y_{\mathrm{P}} = 0.205\pm0.031$ rises to $Y_{\mathrm{P}} = 0.226\pm0.017$ if $n_s=1$. The constraints on the Helium abundance are sensitive to the inclusion of SH0ES when $n_s$ is free, as the Helium fraction shifts to $Y_{\mathrm{P}} = 0.258\pm0.013$ for Planck and $Y_{\mathrm{P}} = 0.242\pm0.029$ for ACT. On the contrary, if $n_s=1$, the constraints remain approximately the same.

\subsubsection{LCDM+$N_{\rm eff}$}

Allowing the effective number of relativistic species $N_{\rm eff}$ to vary provides the best framework for studying the disagreement between Planck and ACT without SH0ES, achieving a tension probability of $p=4.4\%$, the highest among the models examined (see \tab{suspiciousness}). Once SH0ES data are included, however, this probability falls to $p=0.08\%$. Enforcing $n_s=1$ drives the tension above $4\sigma$, both with and without SH0ES.

For Planck and ACT individually, adding $N_{\rm eff}$ does not dramatically affect the Hubble tension unless $n_s$ is fixed. Under an HZ spectrum, Planck exhibits a large decrease, whereas for ACT it remains almost unchanged:
\begin{align*}
    &\qdmap = 4.29\xrightarrow{n_s=1}1.86\quad\text{(Planck)}\\
    &\qdmap = 3.38\xrightarrow{n_s=1}3.99\quad\text{(ACT)}
\end{align*}

Examining the $\Delta$AIC values in \tab{AIC}, this extension of the standard model is generally not favored over $\Lambda$CDM; most cases show at least a \textit{positive} preference for $\Lambda$CDM, except for ACT+SH0ES with $n_s$ free, where the evidence is only \textit{weak}, and Planck+SH0ES ($n_s$ free), which notably shows \textit{weak} evidence in favor of this alternative model, with $\Delta$AIC$=3.59$.

In the HZ spectrum scenario, Planck requires significantly more relativistic degrees of freedom, with $N_{\rm eff}=3.71\pm0.11$ at $68\%$ CL, and this remains stable after adding SH0ES. Conversely, ACT data indicate only a small deviation from the standard expectation value~\cite{ACT:2020gnv}, which actually increases when $n_s=1$, yielding $N_{\rm eff}=2.81\pm0.021$ at $68\%$ CL.

\subsubsection{EDE}

Early Dark Energy proves highly effective at addressing the Planck and ACT tension, consistently yielding tension levels below $2.3\sigma$. EDE with $n_s=1$ achieves the best tension probability among all the extensions considered, while EDE with free $n_s$ is the second-best. Including SH0ES retains a relatively high suspiciousness, but combining SH0ES with an HZ spectrum dramatically reverses the conclusions, with tensions beyond $4\sigma$:
\begin{align*}
    &\log{S}=-4.71\xrightarrow{n_s=1}-4.74\quad\text{(without SH0ES)}\\
    &\log{S}=-4.97\xrightarrow{n_s=1}-12.61\quad\text{(with SH0ES)}
\end{align*}

EDE is also the best model for alleviating the Hubble tension with Planck alone: $\qdmap=2.77$. It remains almost stable when we change the assumption on $n_s$, as it ranges from $2.77$ to $2.29$ when we fix it. For ACT, $\qdmap$ shows an even smaller excursion, with $\qdmap=2.16$ increasing slightly to $\qdmap=2.19$.

Turning to the $\Delta$AIC values in \tab{AIC}, the best performance of this model, as expected, is when SH0ES is combined with Planck. In fact, we obtain very strong evidence whether $n_s$ is a free parameter or fixed, with $\Delta$AIC$=16.05$ and $\Delta$AIC$=13.93$, respectively. Planck-only data do not favor EDE over $\Lambda$CDM, while ACT-only indicates \textit{weak} support when $n_s=1$ and slightly disfavors it when $n_s$ is free. In contrast, for ACT+SH0ES, there is strong evidence for EDE in both cases.

The bound obtained with Planck alone for the fraction of EDE shifts to nonzero values when SH0ES is included. The preference for a nonzero EDE component increases under the HZ scenario, with $f_{\rm EDE}=0.135\pm0.020$ at $68\%$ CL for Planck, and a similar value for Planck+SH0ES. On the other hand, ACT displays the opposite behavior: for ACT+SH0ES, the fraction changes from $f_{\rm EDE}=0.123\pm0.031$ ($n_s$ free) to $f_{\rm EDE}=0.106\pm0.027$ ($n_s=1$), both at $68\%$ CL, with a similar trend in ACT-only analyses.

\section{Conclusions} 
\label{sec:concl}

In this work, we have explored a variety of beyond-$\Lambda$CDM extensions to reconcile Planck and ACT datasets. We include modified geometry ($\Omega_k\neq0$), running of the spectral index ($\alpha_s\neq0$), a general dark energy equation of state ($w\neq-1$), extra relativistic species ($\neff\neq3.044$), a free primordial helium fraction ($Y_p$), also combined with a prior from BBN measurements, and early dark energy. For each case, we also impose a Harrison-Zeldovich spectrum ($n_s=1$) to test its impact on the overall parameter constraints, dataset consistency, and the Hubble tension. 

To quantify the model's performance, we employ multiple statistical tools. We test the overall preference of a fixed dataset combination for an alternative model using the $\Delta$AIC and Bayes factor. Our results align with previous conclusions presented in the literature, indicating that $\Lambda$CDM$+\Omega_k$ and EDE are the preferred models for Planck and Planck+SH0ES, respectively. On the other hand, ACT favors $Y_p$ as a free parameter, while ACT+SH0ES prefers EDE, both combined with an HZ spectrum. The $\Delta$AIC results yield similar, though not identical, preferences, as ACT favors $\Lambda$CDM+BBN, while ACT+SH0ES selects $\Lambda$CDM$+\Omega_k$. This highlights the simplistic penalization of extra parameters in the Akaike criterion, which, while useful for broad comparisons due to its widespread use, does not fully capture the complexity of parameter constraints in cosmological models.

The spectral index is more sensitive to the underlying cosmology, with Planck and ACT agreeing at the $1\sigma$ level when considering extensions that include the running of the spectral index, the helium fraction, the effective number of relativistic species, and early dark energy. However, when combined with SH0ES, these tensions increase. Conversely, $\Omega_bh^2$ is not significantly affected by the inclusion of SH0ES, while imposing a Harrison-Zeldovich spectrum increases the discrepancy between values.

In terms of the tension between Planck and ACT, we note that the lowest tension levels occur for $\Lambda$CDM$+\neff$, EDE, and EDE$+n_s=1$ when SH0ES is not included, and for $w$CDM$+n_s=1$, $w$CDM, and EDE when SH0ES is included. However, the tension probability never exceeds $p=4.4\%$. 

Concerning the Hubble tension, the $\qdmap$ values tend to diminish when the spectral index is fixed to unity in the Planck dataset, with all models presenting a lower value than $\qdmap^{\Lambda\mathrm{CDM}}=5.31$, except for the free-curvature case. ACT does not exhibit major differences with or without an HZ spectrum. 

Overall, we find that no single extension entirely resolves the tensions among Planck, ACT, and SH0ES. From a physical standpoint, these findings underscore that small modifications to the standard model may not suffice to explain the issues observed in cosmological data. Even where statistical evidence marginally favors certain scenarios, significant residual discrepancies remain, raising the question of whether deeper extensions are needed. It is also possible that small systematic effects in CMB data or local $H_0$ determinations are magnified by the precision of modern measurements, highlighting the importance of improved calibration strategies and next-generation surveys.

While CMB measurements are often treated as the cornerstone of precision cosmology, the assumption that their systematics are fully under control may be overly optimistic. Planck, ACT, and SPT each employ different experimental designs, data processing pipelines, and foreground treatments, yet they do not always yield consistent parameter constraints. The Planck-ACT tension, in particular, suggests that residual systematics may still be present, despite extensive efforts to account for them. Given that these datasets play a central role in defining the standard model of cosmology, it is crucial to adopt a more critical approach toward their uncertainties. Moving forward, independent analyses, cross-experiment consistency tests, and novel observational strategies will be essential in determining whether these tensions point to new physics or instead reflect limitations in our current understanding of the CMB.

As this work was being finalized, the new release of ACT-DR6 was announced. This upcoming dataset will provide a new opportunity to reassess the Planck-ACT inconsistency and test whether the tension persists. Whether this discrepancy remains or is alleviated with improved data will be of great interest to the community.

\section*{Acknowledgments}
We thank William Giar\'e for the interesting and useful discussions and for his contribution in developing the Cobaya wrapper for the MCEvidence and the Suspiciousness analysis. MF is funded by the PRIN (Progetti di ricerca di Rilevante Interesse Nazionale) number 2022WJ9J33.
EDV is supported by a Royal Society Dorothy Hodgkin Research Fellowship.
This article is based upon work from the COST Action CA21136 ``Addressing observational tensions in cosmology with systematics and fundamental physics'' (CosmoVerse), supported by COST (European Cooperation in Science and Technology). We acknowledge IT Services at The University of Sheffield for the provision of services for High
Performance Computing.

\bibliography{Bibliography}
\end{document}